\documentclass[useAMS,usenatbib]{mn2e}

\usepackage{mathptmx,subfigure,graphicx,lpic}

\newcommand\aj{{AJ}}%
\newcommand\apj{{ApJ}}%
\newcommand\apjl{{ApJ}}%
\newcommand\apjs{{ApJS}}%
\newcommand\aap{{A\&A}}%
\newcommand\aaps{{A\&AS}}%
\newcommand\baas{{BAAS}}%
\newcommand\mnras{{MNRAS}}%
\newcommand\nat{{Nature}}%
\title[Arc statistics in the Millennium simulation compared to observations]{Lensed arc statistics: comparison of Millennium-simulation
galaxy clusters to Hubble Space Telescope observations of an X-ray selected
 sample}

\author[Horesh et al.]
{Assaf~Horesh,$^{1}$
Dan~Maoz,$^{2}$
Stefan~Hilbert,$^{3,4}$
Matthias~Bartelmann$^{5}$\\
\\
$^{1}$Division of Physics, Mathematics and Astronomy, California
Institute of Technology, Pasadena, CA 91125, USA\\
$^{2}$School of Physics and Astronomy, Tel Aviv University, Tel
Aviv 69978, Israel\\
$^{3}$Argelander-Institut fuer Astronomie, Univresitaet Bonn, 53121
Bonn, Germany\\
$^{4}$Max-Planck-Institut f{\"u}r Astrophysik,
Karl-Schwarzschild-Stra{\ss}e 1, 85741 Garching, Germany\\
$^{5}$Institut fier Theoretische Astrophysik, Universitaet Heidelberg,
69120 Heidelberg, Germany\\}%

\begin{document}

\maketitle
\label{firstpage}
\begin{abstract}
It has been debated for a decade whether
there is a large overabundance of 
strongly lensed arcs in galaxy clusters, compared to expectations from 
$\Lambda$CDM cosmology. We perform ray tracing through the most 
massive halos of the Millennium simulation at several redshifts in their 
evolution, using the Hubble Ultra Deep Field as a source image,
to produce realistic simulated lensed images. 
We compare the lensed arc statistics 
measured from the simulations to those of a sample
of 45 X-ray selected clusters, observed with the Hubble Space Telescope,
that we have analysed in Horesh et al. (2010). The observations and the 
simulations are matched in cluster masses, redshifts,  
observational effects, and the algorithmic arc detection and selection.
At $z=0.6$ there are too few massive-enough clusters in the Millennium
volume for a proper statistical comparison with the observations. 
At redshifts $0.3<z<0.5$, however, we have large numbers of 
simulated and observed clusters, and the latter are an unbiased
selection from a complete sample. For these redshifts, we find
excellent agreement  between the observed and simulated arc
statistics, in terms of the mean number of arcs per cluster,
the distribution of number of arcs per cluster, and the 
angular separation distribution. 
At $z\approx 0.2$ some conflict remains, with 
real clusters being $\sim 3$ times more efficient arc producers than 
their simulated counterparts. This may arise due to selection biases
in the observed subsample at this redshift, to some mismatch in
masses between the observed and simulated clusters, 
or to physical effects that 
arise at low redshift and enhance the lensing efficiency, but which 
are not represented by the simulations.  
\end{abstract}


\section{Introduction}

Galaxy clusters are the largest bound structures in the universe, and
are natural laboratories for studying astrophysical processes and
cosmology. Cluster formation times depend on the cosmological model
(e.g., Richstone, Loeb \& Turner 1992). When a cluster becomes a self gravitating entity, it detaches from the universal
expansion and therefore contains information on the
mean density of the universe at that time. Thus, 
measuring the mass function and mass profiles of
clusters has proved to be important for constraining cosmological
parameters (e.g., Voit 2005; Mortonson, Hu, \& Huterer 2010).
Since the discovery of a gravitationally lensed arcs in galaxy
clusters (Lynds \& Petrosian 1986; Soucail 1987), 
gravitational lensing has been used to study the
evolution of cluster profiles and masses, as well as other cluster
characteristics having potential diagnostic power, such as
ellipticity and substructure.  
One approach has been to perform
detailed modeling of individual clusters using weak and strong lensing
(e.g., Abdelsalam, et al. 1998, Broadhurst et al. 2005, Leonard et al.
2007). However, since this kind of approach is mostly suited for
clusters which exhibit numerous lensed features, the results may
not be representative of the vast majority of clusters. Hence, a
complementary approach is to measure the statistics of lensed arcs in
samples of clusters, studied as a
population.

In an early study of arc statistics, Bartelmann et al.
(1998; hereafter B98), compared the number of giant arcs
in an observed sample to predictions from $\Lambda {\rm CDM}$
cosmology. Using ray tracing, they lensed artificial background
galaxies at $z=1$ through galaxy clusters formed in an N-body
simulation. The lensing cross sections they derived from their lensed images,
together with the density of background galaxies measured by Smail et
al. (1995), were used to predict the number of arcs over the whole
sky. The predicted number of arcs were compared to the number
observed in a sample of 16 X-ray-selected clusters from the Einstein
Observatory Extended Medium Sensitivity Survey (EMSS; Le Fevre et al.
1994). B98 found that the observed number of arcs was higher by an
order of magnitude than the predictions of the (now-standard)
$\Lambda$CDM cosmological model.


The ``order of magnitude'' problem pointed out by B98 stimulated
 several subsequent studies, both observational 
(e.g., Zaritsky \& Gonzalez 2003; Gladders et al. 2003) 
and theoretical.
Wambsganss et al. (2004) studied the dependence of the cross
section for arc formation on the lensed source redshift. They found
that the cross section is a steep function of source redshift, and
concluded that the  problem raised by B98 could be
resolved by adding sources at redshifts greater than $z_{s}=1$ to the
simulations.  Li et al. (2005), however, found that the cross
section dependence on source redshift is shallower than the one found
by Wambsganss et al. (2004). Dalal et al. (2004) repeated the B98 ray
tracing analysis of artificial sources using a larger sample of
simulated clusters (of which B98 had used a subset) and compared their
results to a larger, 38-cluster, EMSS sample (Luppino et al. 1999). They
concluded that the observed arc statistics and the $\Lambda$CDM model
predictions are consistent.

The artificial clusters used by the above studies represented the
cluster dark matter component only. Adding a mass component associated
with the baryonic mass of the cluster galaxies to the artificial
clusters can affect the cross section for forming giant arcs in
several ways (Meneghetti et al. 2000). The critical lines will curve
around the cluster galaxies, thus increasing the cross sections. On the
other hand, the galaxies will split some of the long arcs into shorter
arclets, thus decreasing the giant arc cross section. After studying
these two competing effects, Meneghetti et al. (2000) concluded that
the effect of cluster galaxies on the lensing cross section is
minor, a conclusion also supported by an analytical study by
Flores et al. (2000). Later, Meneghetti et al. (2003) found
that the cD galaxy in each cluster does increase the cross section,
but only by up to $\sim 50\%$ for realistic parameters. Puchwein et
al. (2005) have studied the indirect lensing influence of 
intracluster gas, through its effect of steepening of the dark matter
profile,  which can increase the lensing cross section by a
factor of $\sim 1.5-3$ (but see Rozo et al. 2008). 

Torri et al. (2004) studied the effect of cluster mergers on arc
statistics by following the lensing cross-section of a simulated
cluster with small time steps, as the cluster evolves. They concluded
that, during a merger, the strong-lensing cross section can be
enhanced by an order of magnitude in an optimal projection,
potentially providing yet another contribution towards resolving the
problem reported by B98, if X-ray selected cluster samples are
dominated by merging halos. The effects of cluster mass, ellipticity,
substructure, and triaxiality on the lensing cross section have also
been studied extensively (e.g., Bartelmann et al. 1995, Oguri et al.
2003, Hennawi et al. 2007, Meneghetti et al. 2007).

As summarised above, the properties of the galaxy clusters and the
redshifts of the sources were at the main focus of most studies in the
past decade.  However, the simulations in most of these studies
were oversimplified in several respects. Artificial,
constant surface-density, background galaxies, often at a single redshift,
were used. No observational effects, other than simple flux limits,
were taken into account when considering arc detection.  A more
realistic approach was taken by Horesh et al. (2005; hereafter H05)
who used the Hubble Deep Field (HDF; Williams et al. 1996) 
as a background image in their
lensing simulations, thus lensing real background galaxies with
realistic light profiles, with each galaxy at its photometric
redshift. In addition, in H05 we created realistic lensed images by
adding observational effects such as detector background, cluster galaxy
light, and the Poisson noise that they produce . 
The lens statistics from the simulations were compared to the
statistics of a sample of 10 X-ray-selected clusters at $z=0.2$
observed at high resolution with the Hubble Space Telescope (HST) by
Smith et al. (2005). The
HST sample and  the simulated cluster sample were
 matched in mass and redshift. Lensed arcs were detected in both the real and
the simulated data using an automatic arcfinder that we developed. In
addition to the objectivity this provided, it permitted measuring and
comparing statistics of fainter and shorter arcs than those considered
in previous studies. From a comparison between the number of arcs per
cluster that we found in our simulation to the number in
 the observed sample, we
concluded that the lensed arc-production efficiency of clusters from
the $\Lambda {\rm CDM}$ simulation was, to within errors,
 consistent with observations.
The H05 results, however, were dominated by small number statistics, 
both in the simulations (the limited volume of the GIF 
N-body simulations by Kauffmann et al. 1999 that we used resulted in only five massive
clusters), and in the observed sample of only $10$ clusters. The small angular 
size of the HDF also raised a concern of cosmic variance in the source population.

In the last few years, cosmological N-body simulations with improved resolution and larger volume have become available, including the Millennium simulation (Springel et al. 2005). Clusters from the Millennium simulation were used by Hilbert et al. (2007) to study the strong lensing optical depth dependence on various parameters, e.g., magnification and source redshift. Hilbert et al. (2008) studied the influence of stellar mass on the lensing optical depth of Millennium simulation halos. They found that adding the stellar mass to dark-matter-only halos increases the cross section for the formation of multiple images, but mainly at small radii, i.e., $r<10''$. Puchwein \& Hilbert (2009) have further found that 
line-of-sight structures can increase the cross section for giant arcs, but this is important only in low mass clusters (A similar result was found by Wambganss et al. 2005). However, Hilbert et al. (2007; 2008) did not perform realistic lensing simulations as in H05. 
Lensing cross sections of clusters that include gas,  
from the smooth-particle-hydrodynamics
(SPH) simulation (Gottl\"{o}ber \& Yepes 2007), have been studied recently 
by Meneghetti et al. (2010) and Fedeli et al. (2010).

From the observational point of view, in Horesh et al. (2010; H10) 
we have recently produced a large, empirical arc statistics sample, 
based on HST observations of $\sim 100$ clusters. This cluster sample was large enough to separate into subsamples, according to optical or X-ray selection, and by cluster redshift, and to analyze the arc statistics of each subsample 
separately. We found that X-ray luminous clusters produce $\sim 1$ arc per cluster, but optically selected clusters are five times less efficient. Optically-selected samples
(which have been used in some arc statistics studies) apparently probe 
lower cluster masses than X-ray selected samples, despite the similar optical 
luminosities of the two types of clusters. 
The size of the X-ray-selected H10 sample of arcs 
lowers significantly the Poisson errors in the observational 
results, compared to the sample of H05. We are therefore now in the position to compare the improved observed statistics with improved model predictions, to see whether the arc-statistics problem can be confirmed or resolved.

In this paper, we present a new set of realistic lensing simulations 
produced by using clusters at various redshifts from the Millennium
simulation as lenses, and lensing the Hubble Ultra Deep Field (UDF),
which is several times larger than the HDF. 
The calculations are
tailored to match the observed X-ray-selected sample of H10. 
In $\S2$, we present the sample of simulated clusters. 
The lensing simulation method is described in $\S3$.  
The simulation results are presented in $\S4$, 
with a comparison in \S5 to the observed arc statistics 
 reported in H10. We summarise our conclusions in 
$\S6$.

\section{The Millennium cluster sample}

 In the present work, we produce a large sample of realistic lensed images using
simulated clusters from one of the largest N-body cosmological
simulations, the Millennium simulation (Springel et al. 2005).
The Millennium simulation consists of $N\sim 10^{10}$ particles in a
box of size $500~h^{-1}{\rm Mpc}$. Each particle has a mass of
$m_p=8.6\times10^{8}h^{-1}{\rm M}_{\odot}$. This is a factor $\sim20$ lower than
the mass of the particles in the GIF simulation of Kauffman et al. (1999),
used in the arc statistics studies of  B98 and H05.
Another improvement
in the Millennium simulation is its spatial resolution of
$5~h^{-1}{\rm kpc}$, compared to $25~h^{-1}{\rm kpc}$ in the GIF simulation.  The $\Lambda {\rm CDM}$ cosmological parameters
used in the simulation are $\Omega_{m}=\Omega_{dm}+\Omega_{br}=0.25$,
$\Omega_{br}=0.045$, $\Omega_{\Lambda}=0.75$, $\sigma_{8}=0.9$, $n=1$
and $h=0.73$, where the Hubble constant ${\rm H}_{0}=100~h~{\rm km~s}^{-1}{\rm
  Mpc}^{-1}$. $\Omega_{m}$, $\Omega_{dm}$, and $\Omega_{br}$ are the total matter, dark matter, and baryonic matter densities (in units of the critical closure density), respectively. $\sigma_{8}$ and $n$ are the normalization and the spectral index, respectively, of the cold dark matter power spectrum. These parameters are consistent with the combined analysis of the first year
WMAP data (Spergel et al. 2003) and the 2dFGRS (Colless et al. 2001). The main difference between these parameters and  the parameters of the 7-year WMAP results (Larson et al. 2010) is in the value of $\sigma_{8}$, for which the most recent estimate is $\sigma_{8}=0.801\pm 0.030$. While $\sigma_{8}$ affects the cluster mass function, it is not expected to have a strong effect on the structure of a cluster of given mass (e.g. Fedeli et al. 2008), and hence our main conclusions will likely not be affected by this choice.

A halo catalog of the full Millennium simulation, which is publicly
available\footnote{http://www.mpa-garching.mpg.de/Millennium/}, was
compiled using a friends-of-friends algorithm with a linking length of
$0.2$ (Lemson et al. 2006). We obtained a set of simulated clusters from the Millennium simulation at redshifts $z=0.2, 0.4, 0.6$. The $z=0.2$ snapshot was chosen for comparison with the results of H05. The
other two snapshots were chosen for comparison with the observed sample
presented in H10. 
Since we are interested in the lensing efficiencies of the most
massive clusters in the simulation, we chose those clusters with
masses
${\rm M}_{200}\geq
0.7\times 10^{15} h^{-1} {\rm M}_{\sun}$, which for the adopted Hubble 
parameter
 correspond to $  {\rm M}_{200}\geq 1\times 10^{15} {\rm M}_{\sun}$. 
Here, ${\rm M}_{200}$ is the mass
  enclosed within $r_{200}$, the radius within which the average
density equals 200 times the critical cosmological density at the
observed redshift. The resulting
 simulated cluster sample consists of $14$ clusters at $z=0.2$, and $7$ and $4$ clusters at redshifts $z=0.4, 0.6$, respectively. Figure $1$ shows the mass distributions of the three simulated subsamples.
\begin{figure}
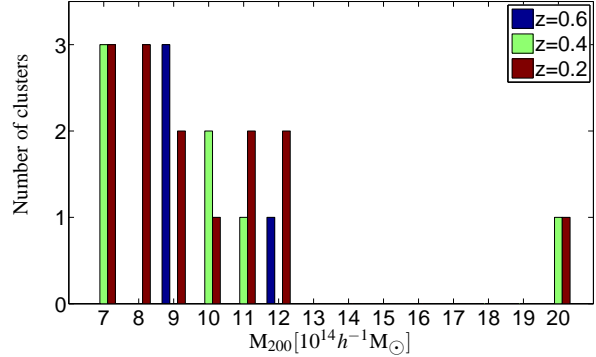

\centering
\begin{lpic}{Millennium_mass_group_bar_plot(9cm)}
\lbl[b]{175,-5;${\rm M}_{200}[10^{14}h^{-1}{\rm M}_{\sun}]$}
\lbl[l]{20,65,90;Number of clusters}
\end{lpic}
\caption{Mass distributions of the simulated Millennium cluster subsamples at $z=0.2$ (red), $z=0.4$ (green), and $z=0.6$ (blue).}
\end{figure}

The deflection angle maps for the above clusters have been produced as described 
in Hilbert et al. (2007). Briefly,
the positions of the particles in each cluster are first projected on
to three orthogonal planes. Then, in order to reduce shot noise, each
particle is smeared into a projected surface mass density cloud of the form
\begin{equation}
\Sigma_{p}(x)=\left\{ 
\begin{array}{ll} 
\frac{3m_{p}}{\pi r_{p}^{2}}\left( 1-\frac{|x-x_{p}|^2}{r_{p}^{2}} \right), & |x-x_{p}|< r_{p}\\
0, & |x-x_{p}|\geq r_{p},
\end{array}  \right \}
\end{equation} 
where $x$ is the position in the lens plane, $x_p$ is the particle
 position, and
 $r_p$ is the three-dimensional distance to the 
64th nearest neighbour particle.
The projected surface mass density profiles for each of
the clusters are converted to two-dimensional
gravitational potentials using the Frigo \& Johnson (2005)
fast-Fourier-transform method. Deflection angle maps are 
produced by differentiating the projected potentials. The maps
are spatially bi-linearly interpolated to achieve a resolution of
$0\farcs03$, which is the resolution of the UDF (Beckwith et al. 2006) 
image that we use as a source image in our lensing simulations 
(see $\S3$, below).

In addition to the dark matter halo catalog, Lemson et al. (2006) have released a public galaxy population catalog. The galaxy properties are based on semi-analytical models by De Lucia \& Blaizot (2007), and Bower et al. (2006). In our work we use the galaxy catalog based on De Lucia \& Blaizot. As in Hilbert et al. (2008) we add the contribution of the stellar mass of the cluster galaxies to the final deflection angle maps. When
 we perform our lensing simulations (see $\S3$), the above catalog is also used for adding the light of the cluster galaxies to our simulated lensed images.   

\section{Lensing simulations}

Our lensing simulations are calculated in a manner similar to that in H05. They are realistic in the sense that we lens real galaxies, pixel by pixel, using a real image of a field of galaxies as the background image in our simulations. We thus incorporate galaxy properties such as the galaxy luminosity function, the redshift distribution, and the size and shape distributions directly into the simulations, avoiding the need to add by hand 
the effects of these inputs later on. 
While in H05 we used the HDF as a source image, in our new simulations we use the UDF as our background source image. 

The UDF
is a  $\sim 1$ Ms exposure of $11$ arcmin$^{2}$ in the southern
sky, obtained using the Advanced Camera for Surveys
 on HST. The exposure time
 was divided among four filters, F435W, F606W, F775W, and F850LP.
The limiting magnitude is $m_{AB}\sim 29$ for point
sources in all bands. Coe et al. (2006) have produced a photometric
redshift catalog of the UDF, containing 8042 objects detected at the
10$\sigma$ level. We use this catalog to build the redshift image
of the UDF, which is necessary for our lensing simulations, in which the pixels
of each source are assigned the source photometric redshift. In addition to that redshift image, we use the UDF F775W filter ($I$ band) image as the source background flux image in our simulations.
In what follows, we use a limiting magnitude of 24 for the detection of simulated
arcs, and hence arcs with  magnifications of 
up to 30 (which are the largest ones we find in practice)  originate from unlensed
sources of $\sim 27.5$ mag. This is still a factor 10 brighter than the limiting magnitude of the 
UDF. Stated differently, the UDF is a sufficiently deep source image for
magnifications up to $\sim 100$, and
such large magnifications behind clusters are unlikely.

We simulated the lensing of the UDF by the artificial 
Millennium clusters by ray tracing,
as follows. A light ray is shot backwards from the observer to a specific pixel position $(i,j)$ in the lens plane of a cluster. The scaled deflection angle at the ray position in the lens plane, $\vec{\alpha}_{i,j}$, is used to deflect the ray backwards. For any single source redshift, the light ray will reach a certain point in the background source image. However, our use of a range of source
redshifts, rather a single one, means that the light ray is actually deflected to multiple positions which translate to a line segment in
 the background source image. 
We therefore search for background galaxies in the source image along this line segment. Each pixel which belongs to one of these galaxies and through which the light ray line passes is then checked individually. The redshift of the galaxy to which the pixel belongs, $z_s$, is plugged into the lens equation 
\begin{equation} 
\vec{\beta}=\vec{\theta}-\vec{\alpha}(\vec{\theta},z_{s}),
\end{equation}
 where $\vec{\theta}$ is the image position, 
 giving the position in the source plane $\vec{\beta}(i,j,z_{s})$ 
to which the light ray is actually deflected for that specific source redshift
(we note that surface brightness is conserved by this mapping). In the case that $\vec{\beta}(i,j,z_{s})$ matches the position of the source pixel which is being examined, that pixel is lensed to the pixel at position $(i,j)$ in the lens plane. By repeating this process for every pixel in the lens plane, we build a lensed image of the UDF. 

In order to simulate the real observed images whose arc statistics we will test,
 we add observational effects to the simulated images. 
These effects include the light of cluster galaxies, backgrounds, photon noise, and readout noise. As mentioned in $\S2$, we use the Millennium simulation galaxy catalog (Lemson et al. 2006) to add the galaxy cluster light to each cluster. The light is added at the positions of the mass overdensities which correspond to galaxies in the above catalog. The light of each galaxy is added in the form of a projected S\'{e}rsic light profile, 
\begin{equation}
\log \left( \frac{I}{I_{e}} \right)=-b_{n}\left[\left( \frac{R}{R_{e}} \right) ^{1/n} -1\right],
\end{equation} 
where $I$ is the surface brightness, $R$ is the radius, 
$R_{e}$ is the effective half light radius, and $I_{e}$ is the surface brightness at $R_{e}$, with a random axis ratio in the range $[0.5,1]$. The profile parameters are randomly drawn from observed distributions. Specifically, the effective light radius is calculated according to the relation between $R_e$ and $i$-band luminosity found by Bernardi et al. (2003), who studied early-type galaxies in the Sloan Digital Sky Survey. The other profile parameters are taken from Caon et al. (1993), 
\begin{equation}
\log ~n=0.28+0.52\log R_{e} ,
\end{equation}
where $R_{e}$ is in kiloparsecs and $b_{n}$ is coupled to $n$ such that $R_{e}$ contains half of the total flux.

Once the light of the cluster galaxies is added, Poisson photon noise is calculated for each pixel based on its total light (lensed galaxies, cluster galaxies, and background). 
The background value is randomly chosen from a Gaussian distribution with a mean of 120 $e^{-}$ per original sized ACS pixel for a total exposure time of $1440~s$, and a $1\sigma$ dispersion of $40~e^{-}$, which is the distribution
 of the background levels that we find in the observed sample of H10. Furthermore, a readout noise of $5 e^{-}$ per exposure is added to the simulated lensed images. 
\begin{figure}
\centering
\includegraphics[width=9cm]{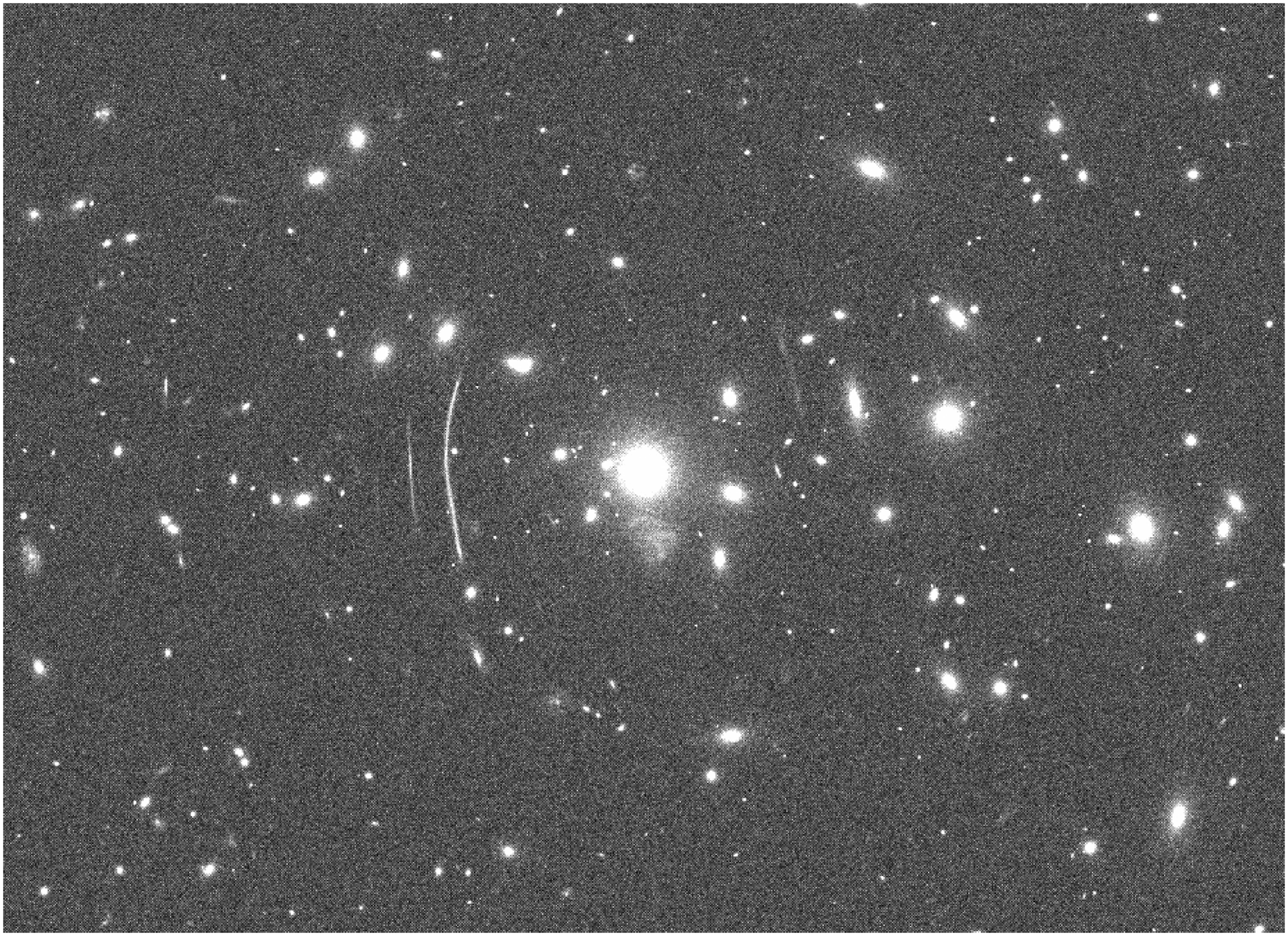}
\vskip 0.2cm
\includegraphics[width=9cm]{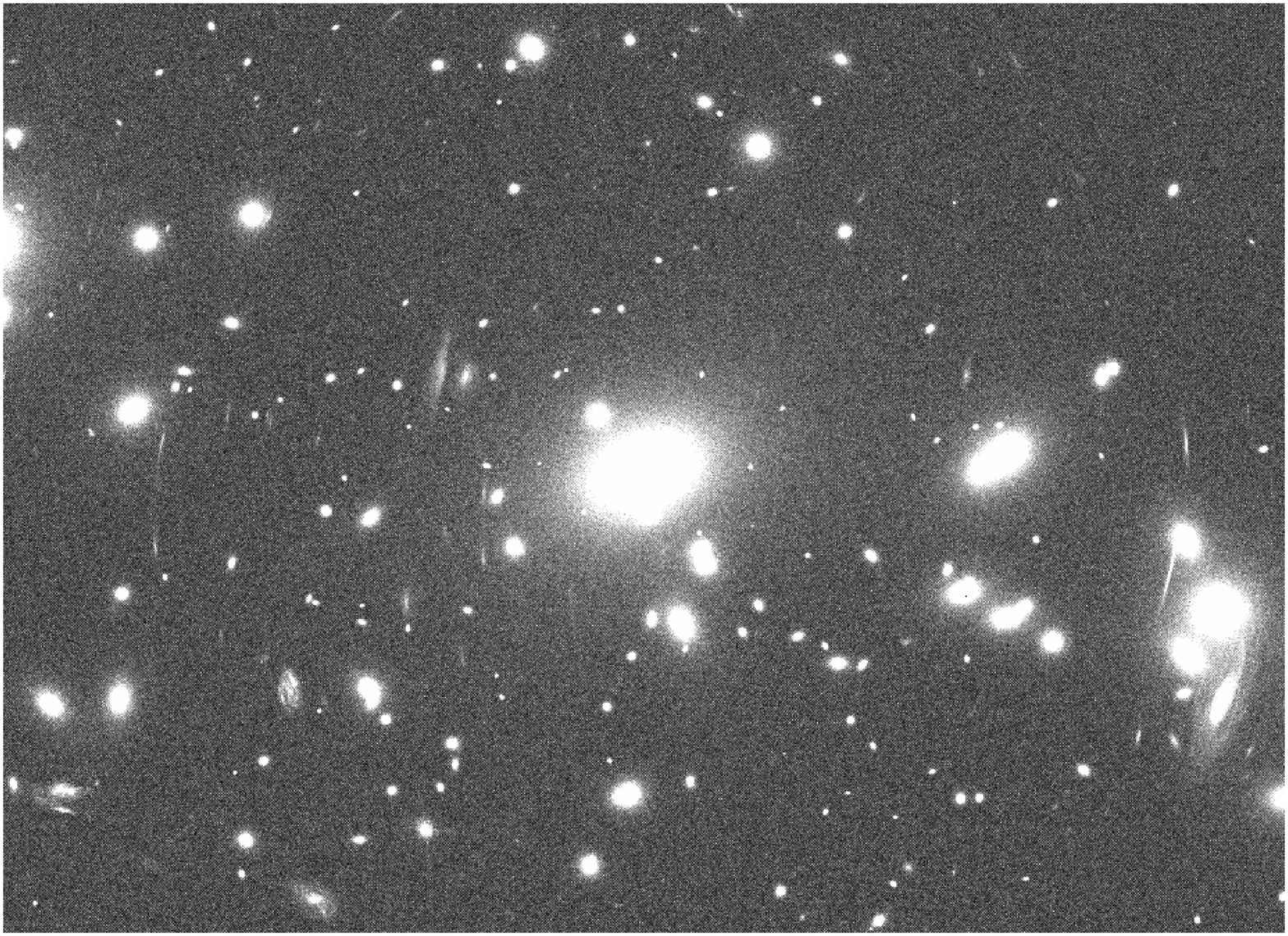}
\caption{Two examples of realistic simulated lensed images. The two
  $160''\times100''$ images are of the same cluster at redshift
  $z=0.4$ (top panel), and 
after it has evolved to $z=0.2$ (bottom panel).} 
\end{figure}

Our simulations consist of three projections and five source background realizations for each cluster, where in each realization the UDF source image is shifted to a different random position behind the lens. Hence, the simulations result
 in $210, 105$, and $60$ simulated images of clusters at redshifts $0.2, 0.4$, and $0.6$ respectively. Figure $2$ shows an example of the simulated images 
of the same cluster, with different realizations, at two different redshift 
snapshots.   

After producing the final set of simulated images, we automatically detect arcs in them. 
In H10, we used two different arc finders
on the observational data, that of H05 and that of Seidel et al. (2007).
In the current analysis of this large simulated dataset, 
to save effort we use only the H05 arc
finder. However, from
our experience in H10, the fraction of additional arcs that we
would have found using both the H05 and Seidel et al. arcfinders, 
compared to the H05
arcfinder alone, is $5\%$ at most, and therefore of little consequence
to our results.

The H05 arc-detection algorithm is based
on application of the SExtractor (Bertin \& Arnouts 1996) object
identification software. The output of repeated SExtractor calls, using
different detection parameters each time, is
filtered using some threshold of object elongation. The final
SExtractor call is executed on an image combined from the filtered
``segmentation image'' outputs of the previous calls.  The arc
candidates detected in that last call are included in the final arc
catalogue if they meet the required detection parameters defined by the
user. 
We apply the same detection thresholds of arc length-to-width ratio
$l/w\geq 8$ and total magnitude $m_{I}<24$,
 that were used  to detect arcs in the observed samples in H10, 
also maintaining a $60''$ search radius around the cluster center. The arcs that are automatically detected are then visually inspected in order to remove spurious detections, such as diffraction spikes and spiral galaxy arms.

\section{Results}

In the simulations described above, we detect, in the 
cluster samples at  $z=0.2, 0.4$, and $0.6$, numbers of arcs
of  $N_{\rm arcs}=137, 90$, and $25$ with $l/w>8$, and $N_{\rm arcs}=84, 64$, and $18$ arcs with $l/w>10$, respectively 
(see Figure $3$ for some examples). 
The arcs span a magnitude range of $18.8<m_{I}<24$, and their $l/w$ ratios are
 in the range $8-44$, with a median value of $14$. As shown in Figure $4$, at least $50 \%$ of the effective simulated cluster lenses, i.e. clusters that produce at least one arc, produce multiple arcs.
\begin{figure}
\centering
\fbox{
\includegraphics[width=3cm, angle=0]{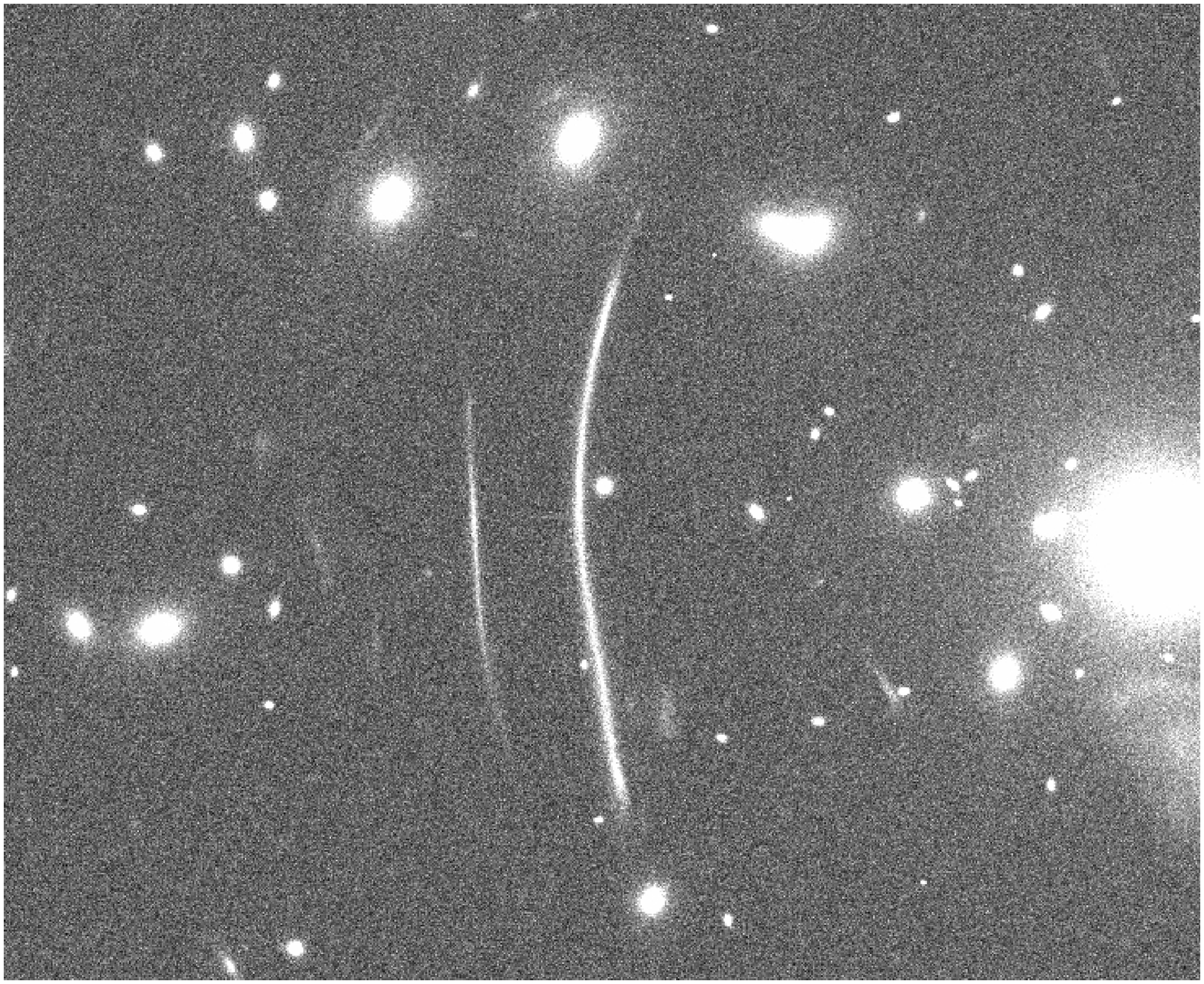}
\includegraphics[width=3cm, angle=0]{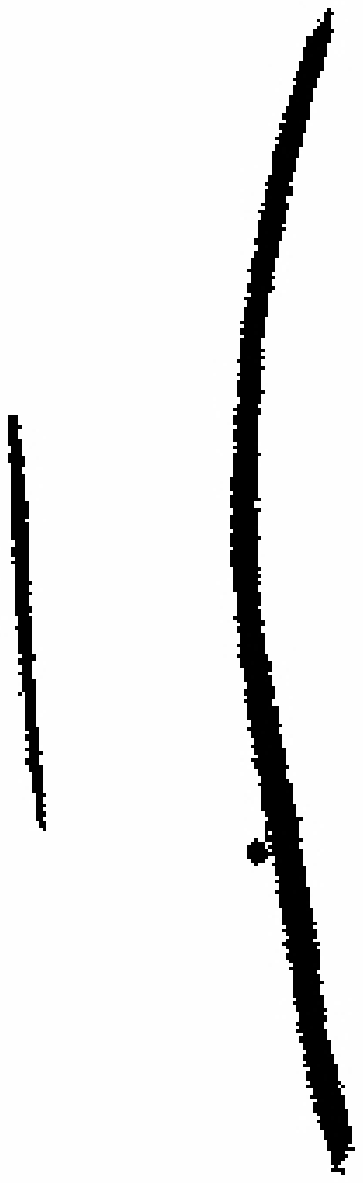}}
\hspace{0.5cm}
\fbox{
\includegraphics[width=3cm, angle=0]{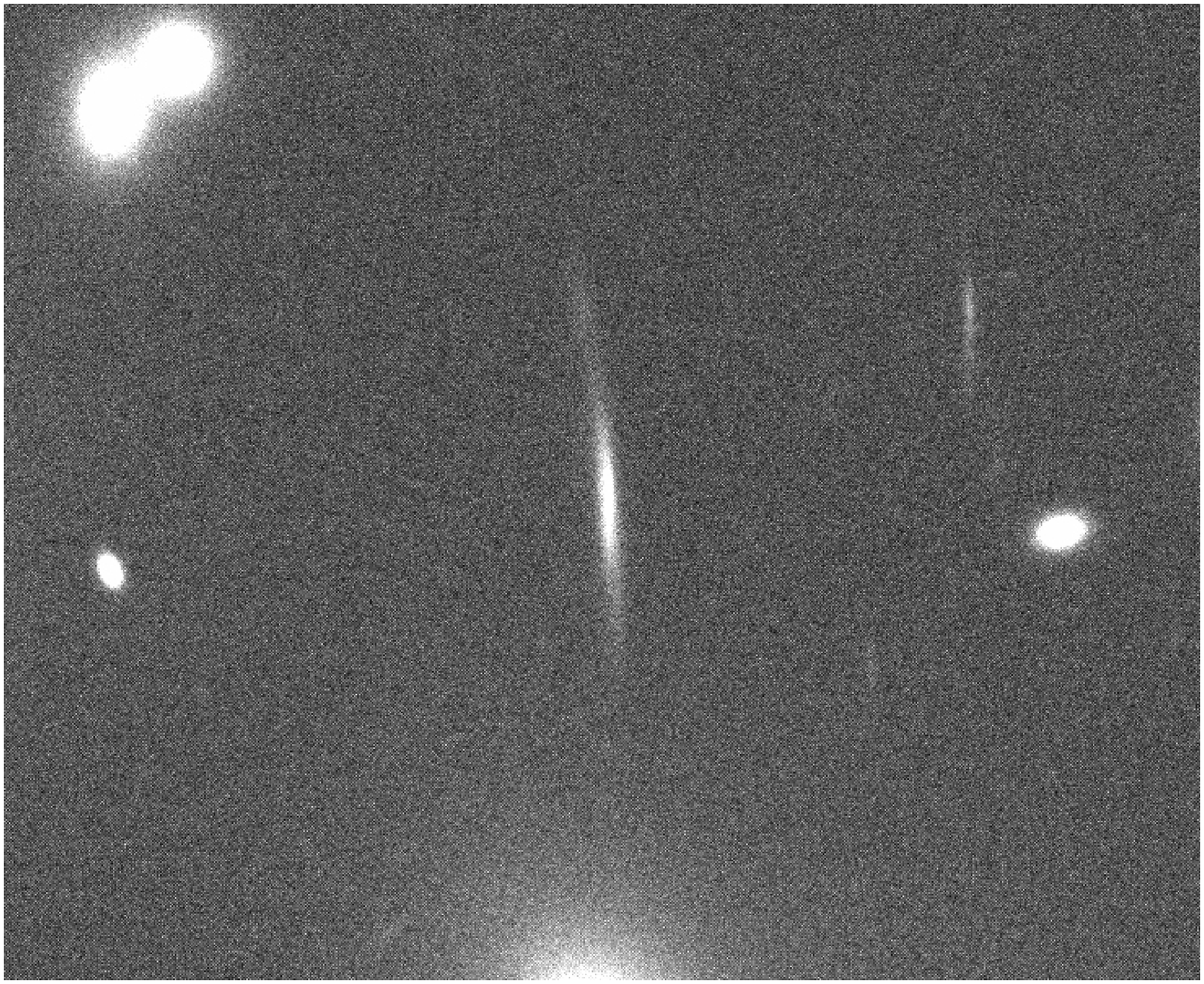}
\includegraphics[width=3cm, angle=0]{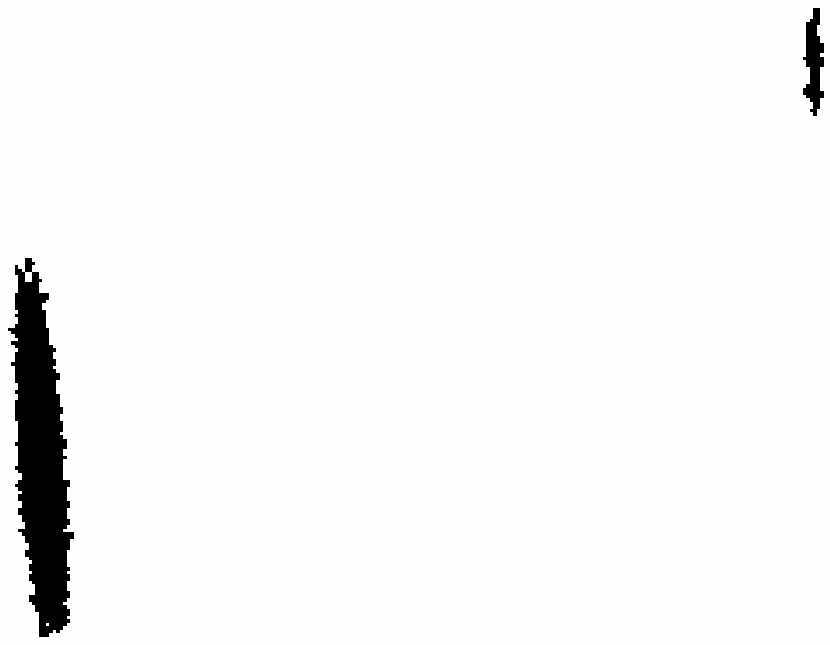}}
\vskip 0.2cm
\fbox{
\includegraphics[width=3cm, angle=0]{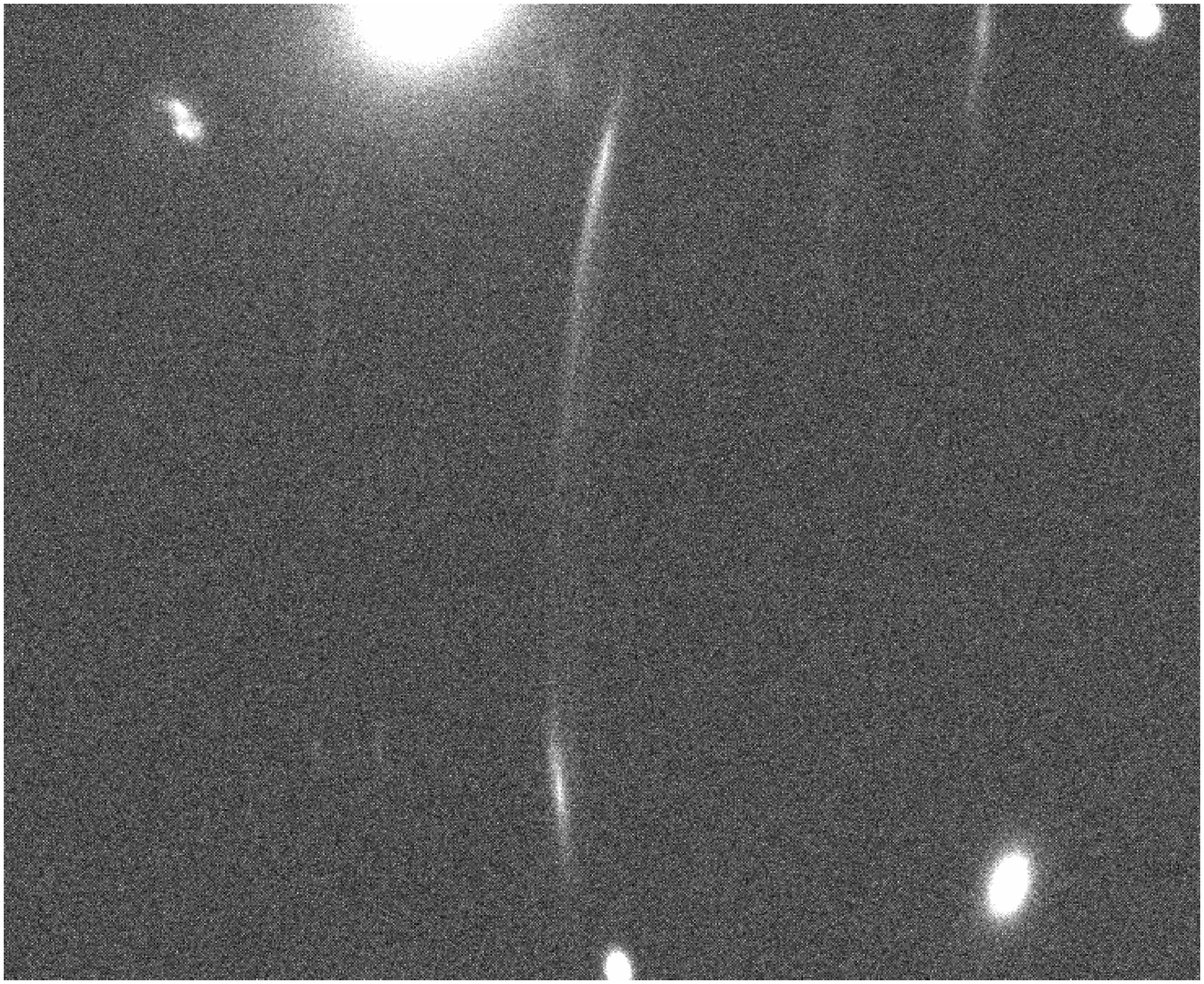}
\includegraphics[width=3cm, angle=0]{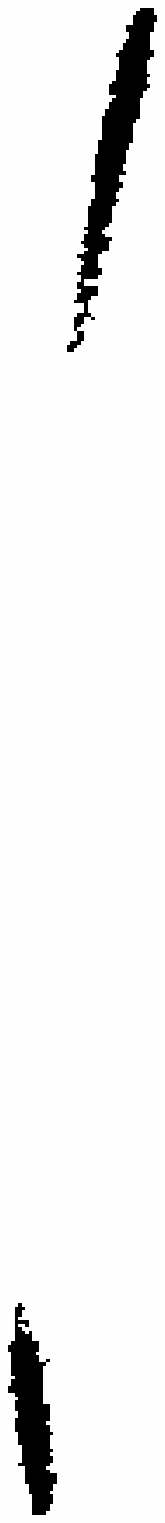}}
\hspace{0.5cm}
\fbox{
\includegraphics[width=3cm, angle=0]{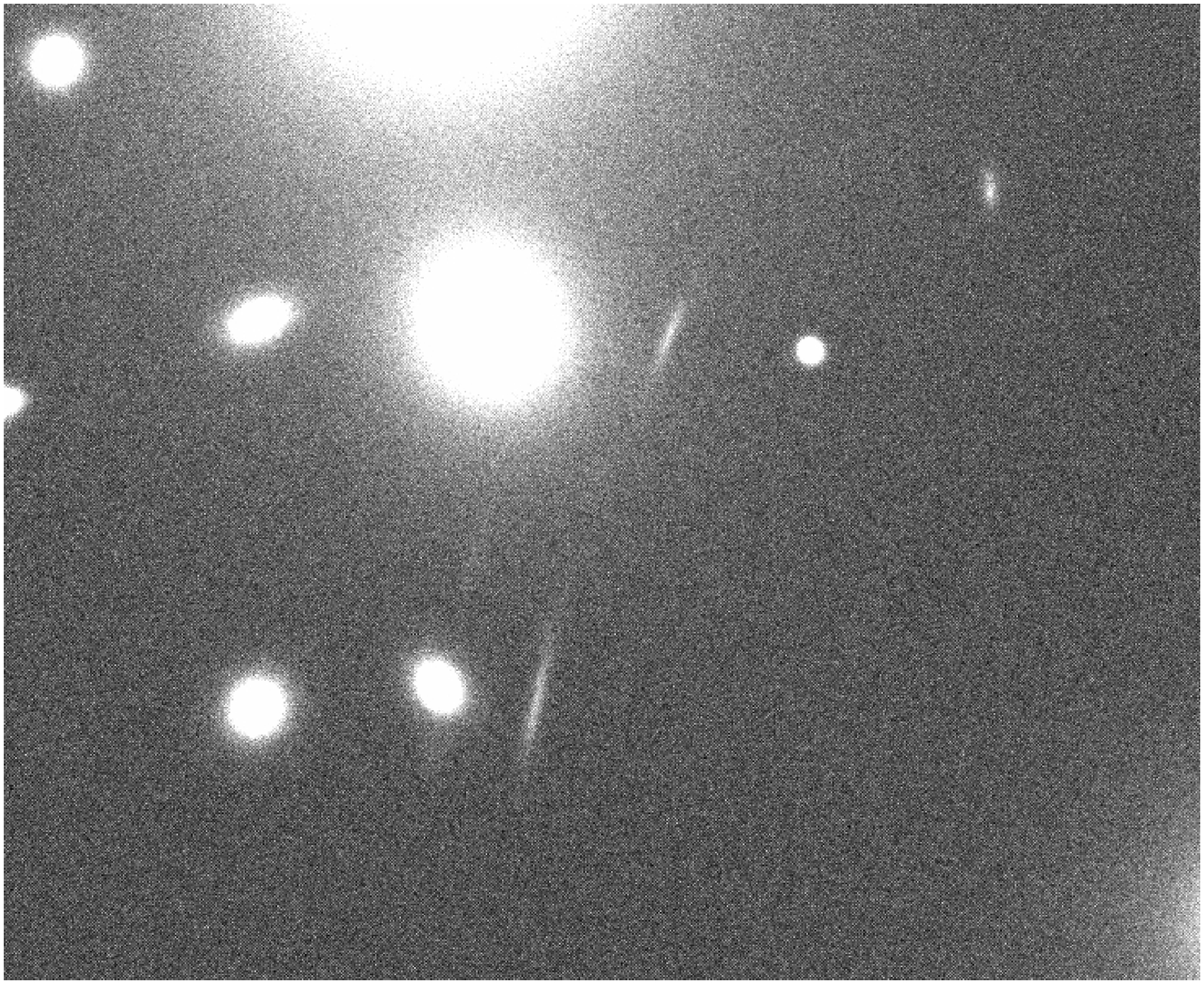}
\includegraphics[width=3cm, angle=0]{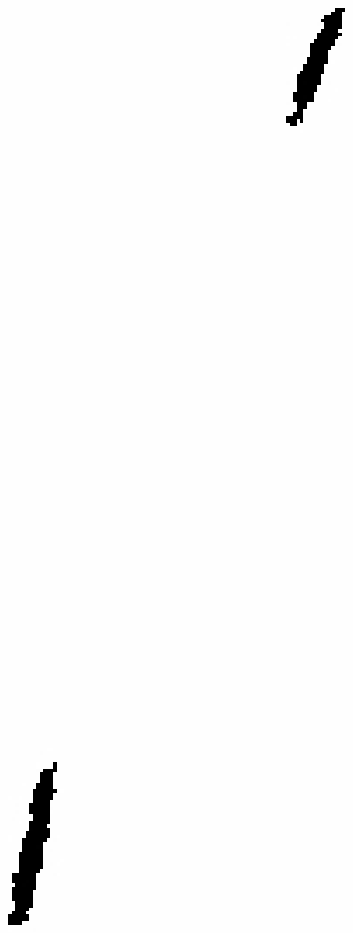}}
\caption{Four examples of simulated arcs (left half of each panel) 
and their detections by the arcfinder algorithm (right halves).}
\end{figure}

\begin{figure}
\centering
\includegraphics[width=9cm]{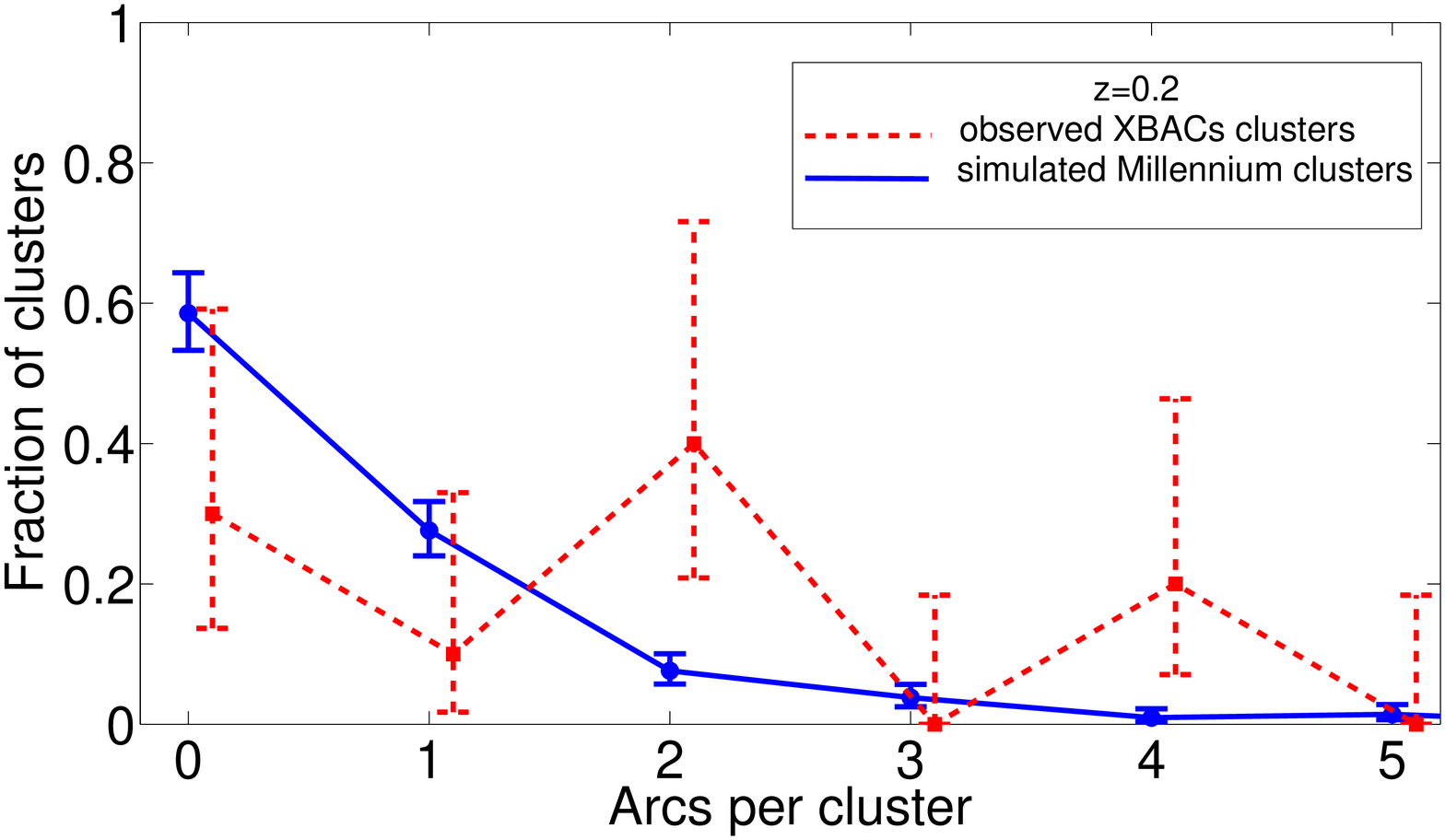}
\includegraphics[width=9cm]{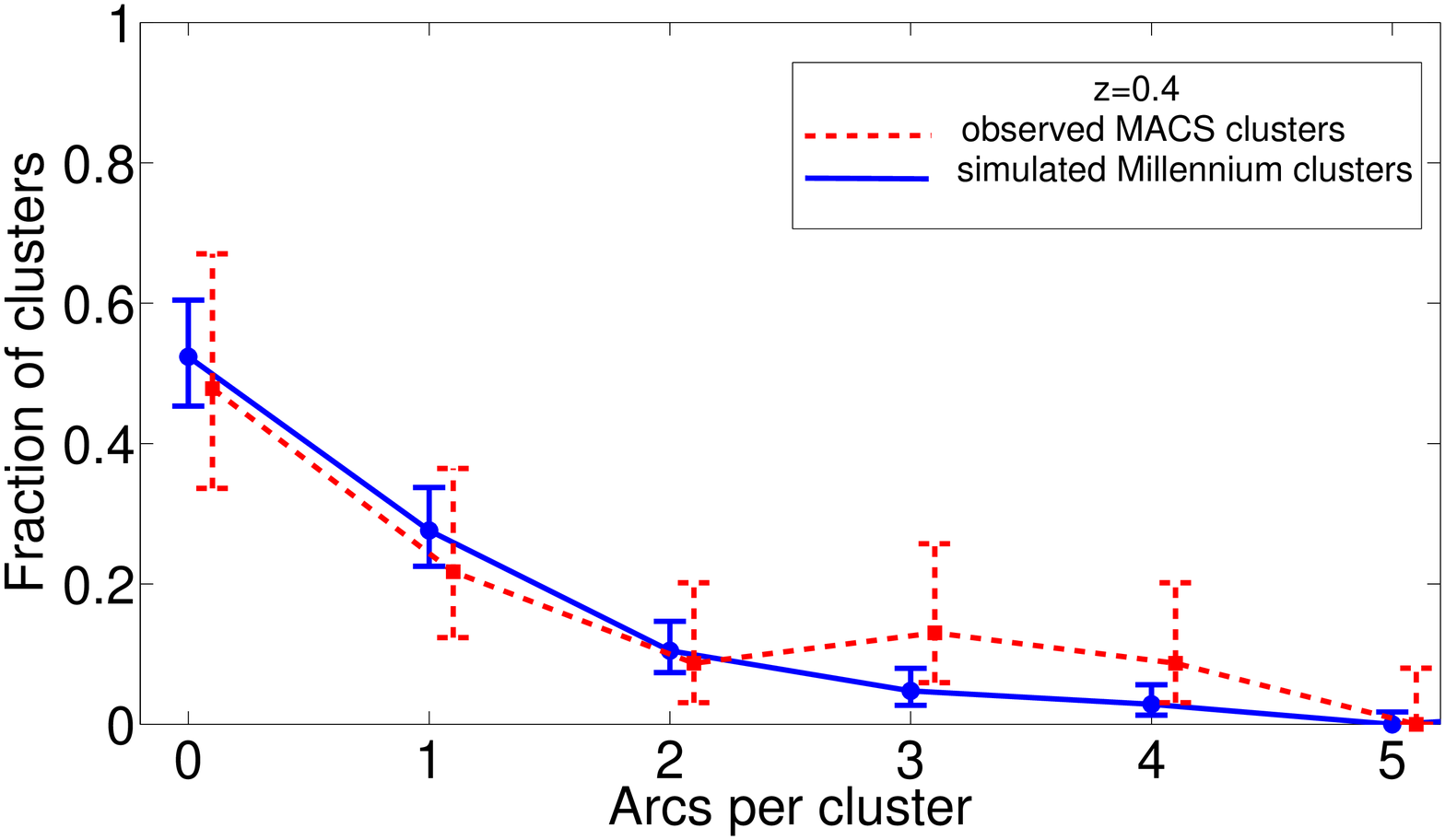}
\includegraphics[width=9cm]{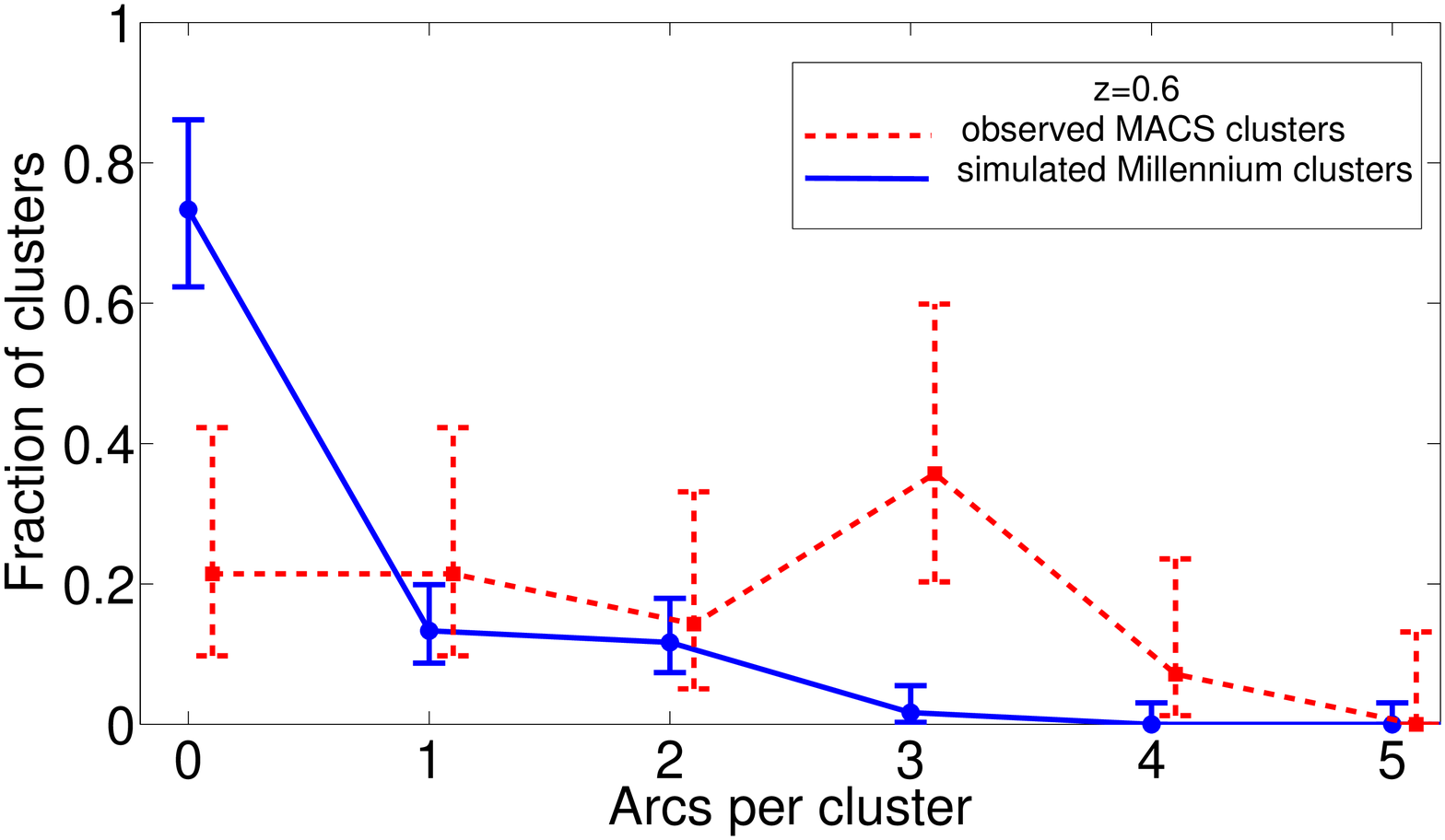}
\caption{Number distributions of arcs in the simulated 
Millennium clusters (blue solid line) and in the observed cluster samples 
(red dashed line): XBACs clusters (H05) and MACS clusters (H10) at 
redshifts $z=0.2$ (top panel), $z=0.4$ (middle panel), and $z=0.6$ 
(bottom panel). Note the excellent agreement, for the $z\approx 0.4$ samples, between the observed and the simulated arc statistics.}
\end{figure}


We now calculate the mean arc production efficiency  -- the number or
detected arcs per cluster --
at each redshift. 
We study the efficiencies of our entire sample and also of the subset that
 is composed of simulated clusters with masses ${\rm M}_{200}\geq
 10^{15}~h^{-1}{\rm M}_{\sun}$. This subset represents the mass range
 of MACS clusters analyzed in H10, and consists of $6, 4$, and $1$
 clusters at  $z=0.2, 0.4$, and $0.6$, respectively (out of the
 original 14, 7, and 4 clusters at these redshifts). It appears that,
 in the entire sample of clusters, the efficiencies, which are
 summarized in Table $1$, peak at
 $z=0.4$, with a $2\sigma$ significance. At that redshift, the entire sample of seven clusters
 produces $0.86^{+0.10}_{-0.09}$ and $0.61^{+0.09}_{-0.08}$ arcs per
 cluster with $l/w\geq 8$ and $l/w\geq 10$, respectively. The subset
 of the most massive clusters produces $1.17^{+0.10}_{-0.09}$ and
 $0.85^{+0.14}_{-0.12}$ arcs per cluster for the two values of $l/w$, 
respectively. Fig. $4$ shows the distribution of arc numbers per cluster, which are compared in \S5, below, to the observed distributions of 
H10.  The errors above, and thoughout the paper, are Poisson.

\begin{table*}
\caption{Arc statistics simulation summary}
\smallskip
\begin{tabular}{lcccccc} 
\hline
\noalign{\smallskip}
Subsample & $N_{images}$ &  $N_{lenses}$ & \multicolumn{2}{c}{$N_{arcs}$} &
\multicolumn{2}{c}{Arcs per cluster}  \\
 & & & ($l/w \geq 8$) & ($l/w \geq 10$) &  ($l/w \geq 8$) & ($l/w \geq 10$)  \\
\noalign{\smallskip}
\hline
\multicolumn{6}{c}{Clusters with mass ${\rm M}_{200} \geq 7\times 10^{14} h^{-1} {\rm M}_{\sun}$}\\
\hline
\noalign{\smallskip}
 $z=0.2$ & 210 & 87 & 137 & 84 & $0.65^{+0.06}_{-0.06}$ &
 $0.40^{+0.05}_{-0.04}$  \\
\noalign{\smallskip}
 $z=0.4$ & 105 & 50 & 90 & 64 & $0.86^{+0.10}_{-0.09}$ &
 $0.61^{+0.09}_{-0.08}$  \\
\noalign{\smallskip}
 $z=0.6$ & 60 & 16 & 25 & 18 &
 $0.42^{+0.1}_{-0.08}$ & $0.30^{+0.09}_{-0.07}$  \\
\noalign{\smallskip}
\hline
\multicolumn{6}{c}{Clusters with mass ${\rm M}_{200} \geq 10^{15} h^{-1} {\rm M}_{\sun}$}\\
\hline
\noalign{\smallskip}
 $z=0.2$ & 90 & 56 & 99 & 59 & $1.1^{+0.11}_{-0.11}$ &
 $0.66^{+0.10}_{-0.09}$  \\
\noalign{\smallskip}
 $z=0.4$ & 60 & 35 & 70 & 51 & $1.17^{+0.10}_{-0.09}$ &
 $0.85^{+0.14}_{-0.12}$  \\
\noalign{\smallskip} 
 $z=0.6$ & 15 & 8 & 12 & 8 &
 $0.80^{+0.40}_{-0.23}$ & $0.53^{+0.26}_{-0.18}$  \\
\noalign{\smallskip}
\hline
\smallskip
\end{tabular} 

Note - The number of images in the second column is calculated by multiplying the number of clusters by the number of projections (3) and by then number of source background realizations (5). The number of lenses in the third column is the number of images in which at least one arc with $l/w\geq 8$ was detected.

\end{table*}

The simulated cluster lenses exhibit some 
arcs at large angular separations from the cluster centers. As shown in Figure $5$, the distributions of arc angular separation are broad. 
The arcs produced by clusters at $z=0.6$ have a median radial separation of $26''$, compared to a median separation of $33''$ in the two lower-redshift snapshots, with a standard deviation of $14''$ in all three redshift snapshots. The angular-separation distributions are compared to the observed distributions of H10 in $\S5$, below.
\begin{figure}
\centering
\includegraphics[width=9cm]{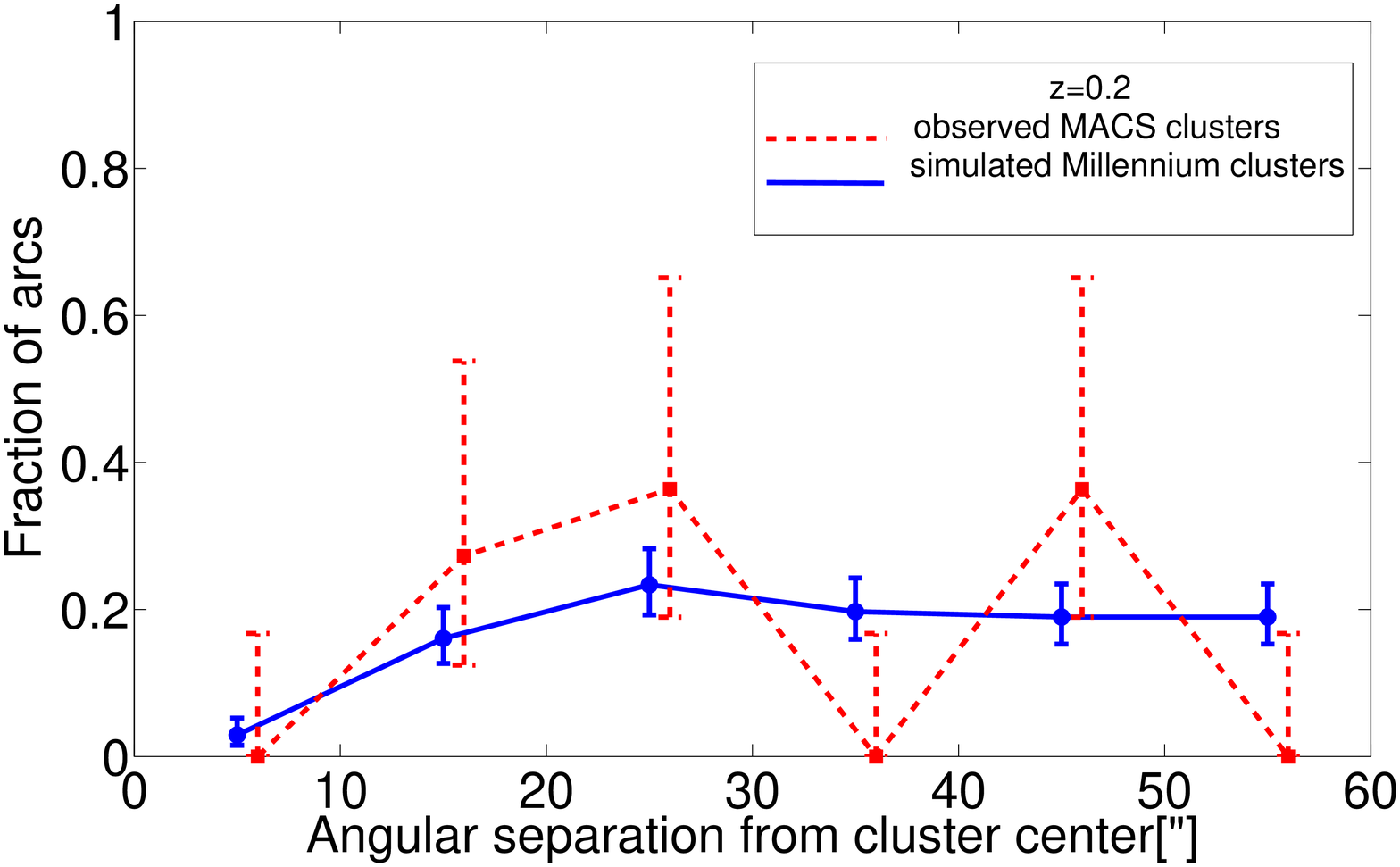}
\includegraphics[width=9cm]{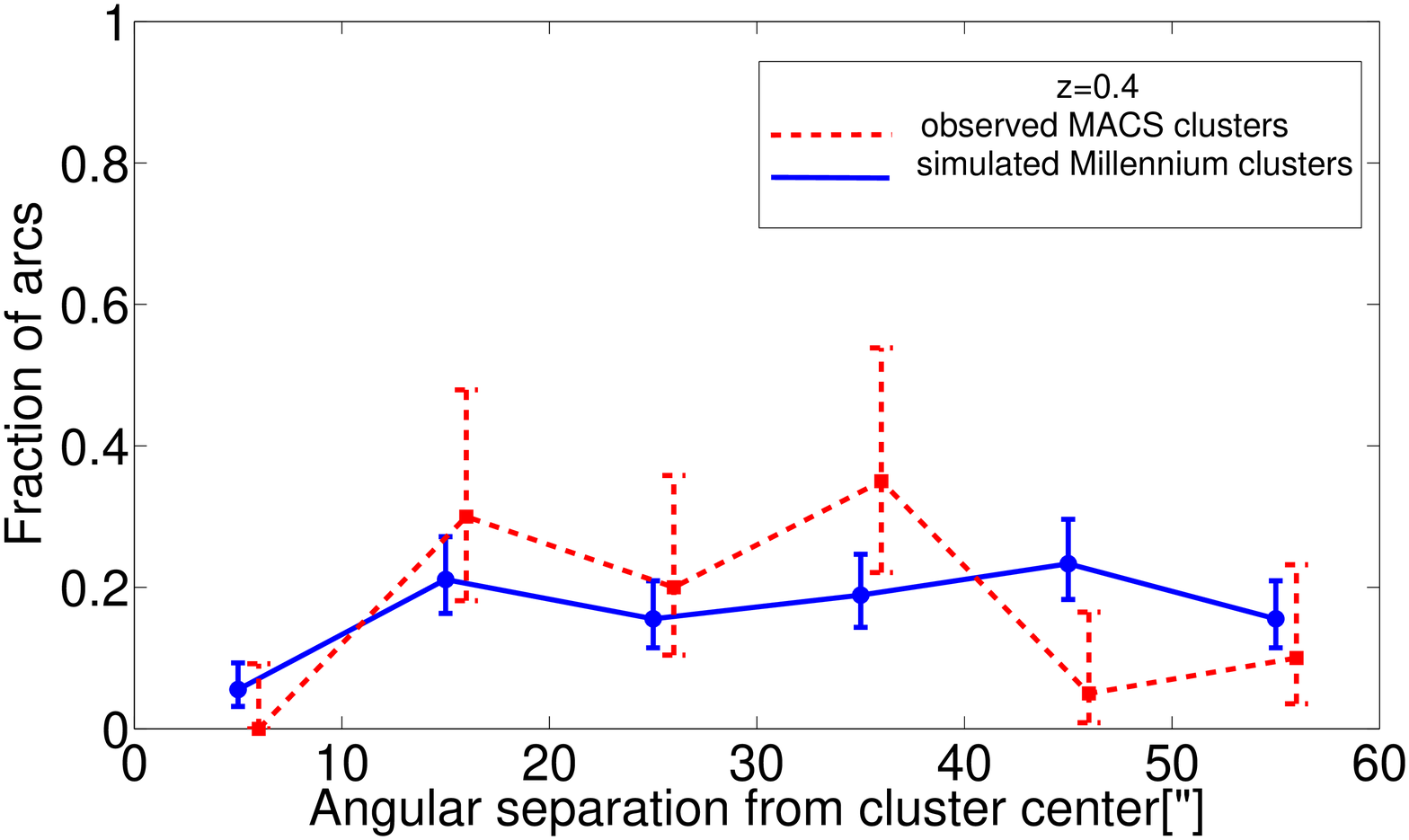}
\includegraphics[width=9cm]{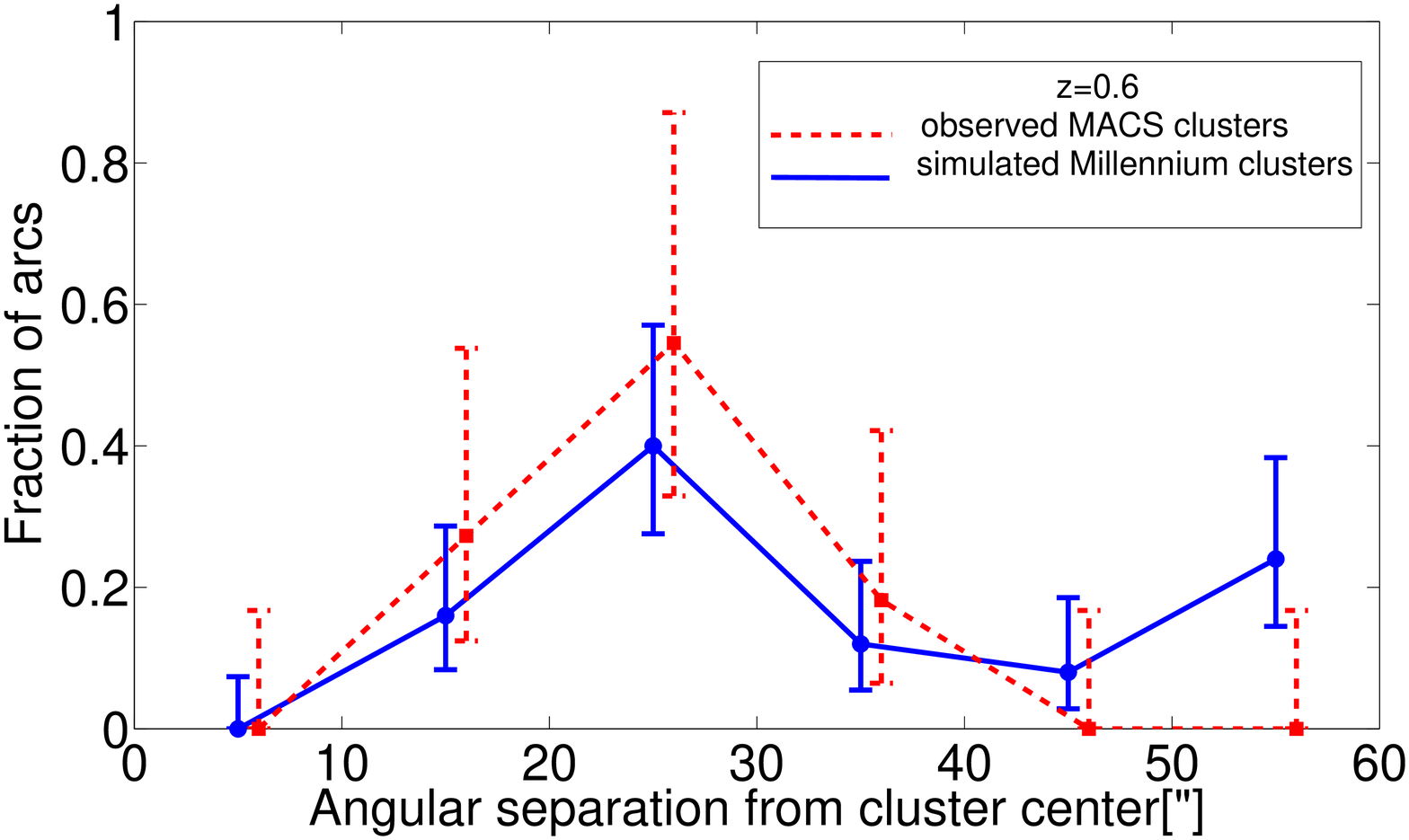}
\caption{As in Fig.~4, but for the distributions 
of arc separations from cluster centers.}
\end{figure}

\subsection{Lensed source properties}

Our simulations allow us also to trace back the background galaxies
 that were lensed into the
 arcs that we detect. We can thus study the properties of this galaxy population in comparison to the full UDF galaxy sample. We first examine the redshift distribution of the lensed arcs. As seen in Figure $6$, the arcs formed by clusters at redshifts $z=0.2$ and $z=0.4$ originate mostly from galaxies at $z\sim (0.5-1.5)$, with a median redshift of $z=1.1$, while the redshift distribution of arcs produced by clusters at $z=0.6$ has a median of $z=2.2$. Our numerically predicted arc redshift
distributions are another statistic that can be compared to ongoing and future observations (e.g. Bayliss et al. 2010).

\begin{figure}
\centering
\includegraphics[width=9cm]{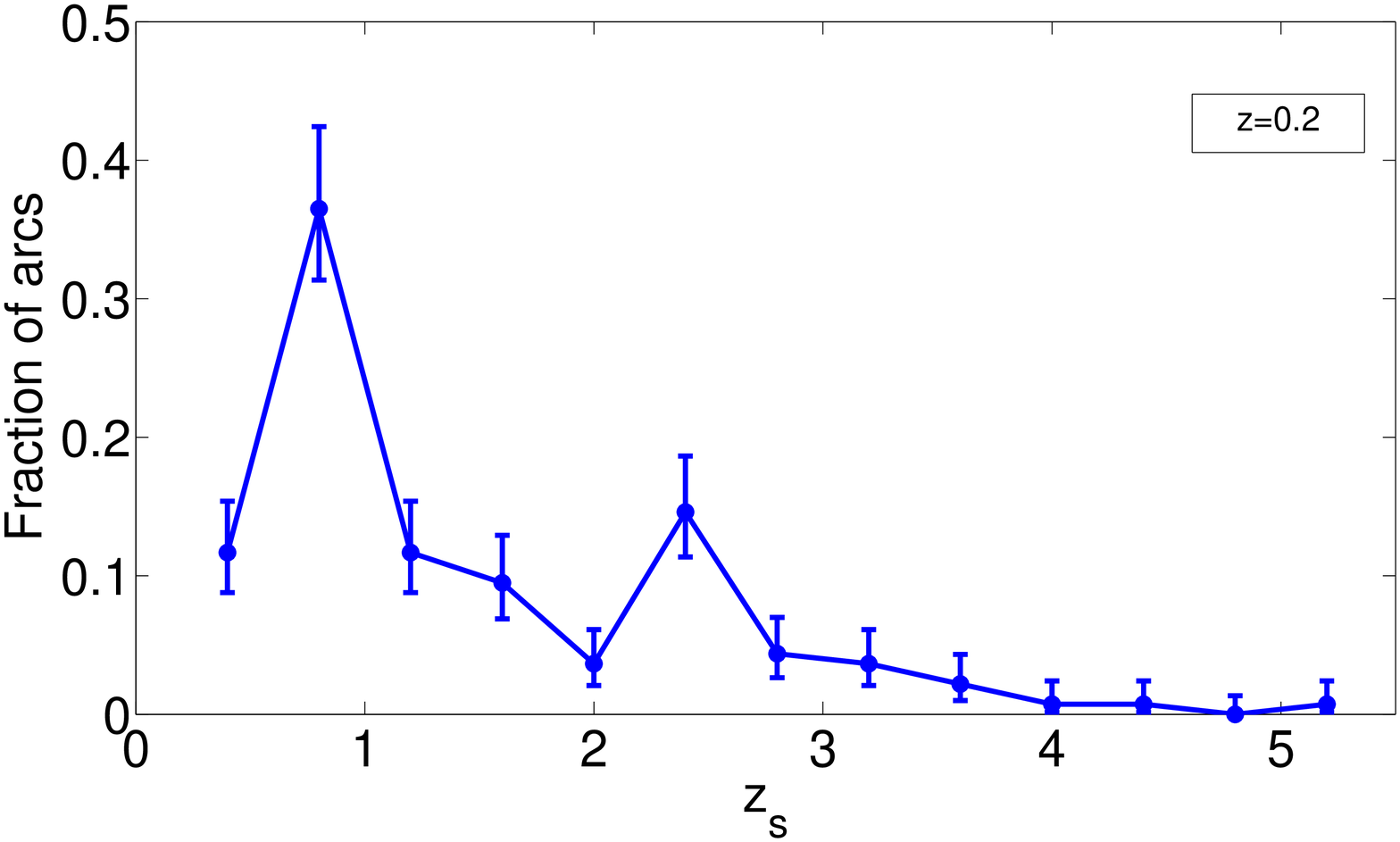}
\includegraphics[width=9cm]{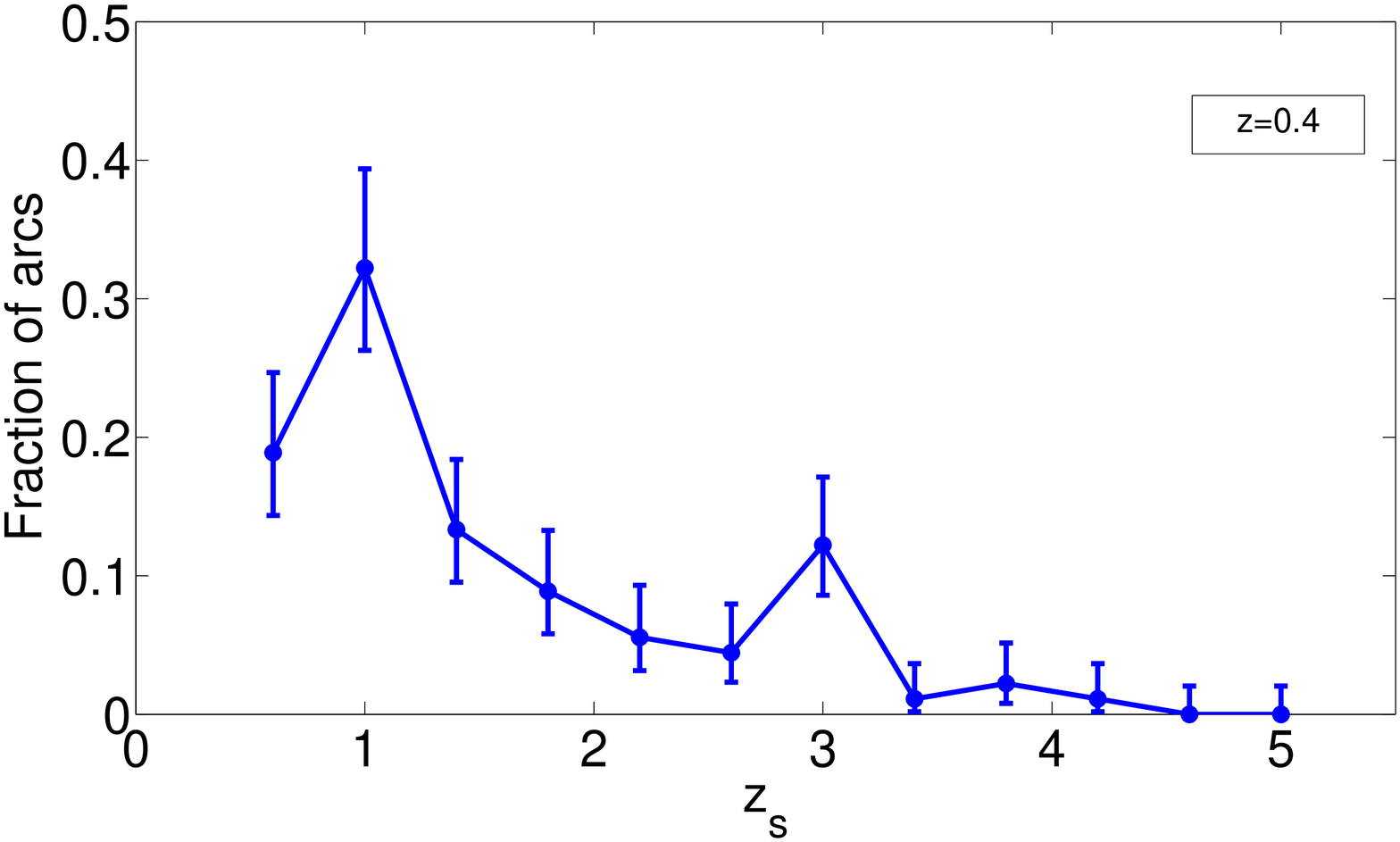}
\includegraphics[width=9cm]{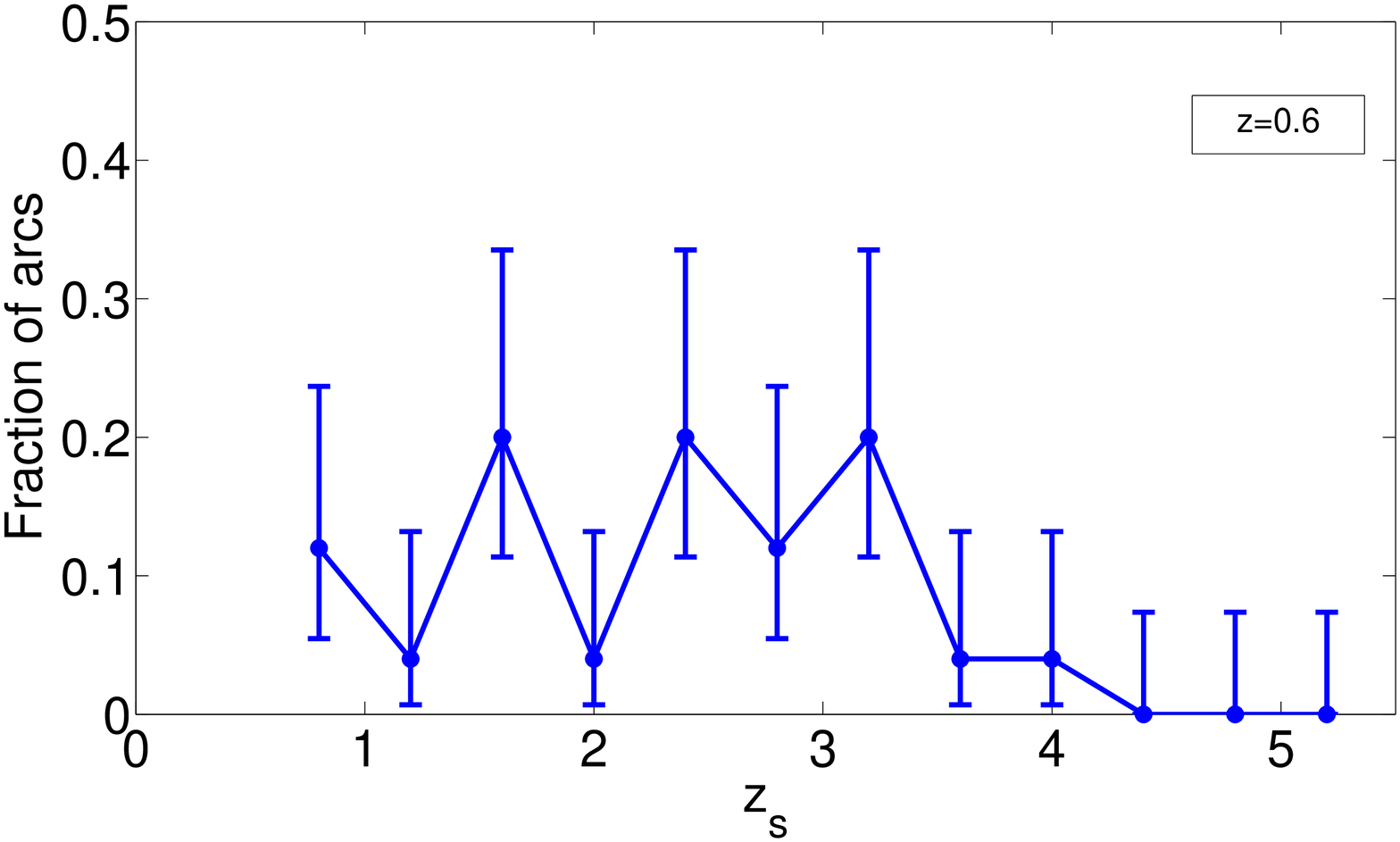}
\caption{Arc redshift distributions in Millennium clusters at $z=0.2$ (top panel), $z=0.4$ (middle panel), and $z=0.6$ (bottom panel).}
\end{figure}

Yet another interesting statistic to look at is the ellipticity distribution of the background galaxies that are lensed into arcs. We have used the Coe et al. (2006) UDF detection image in order to fit an ellipse to each UDF galaxy. The measurement of the semi-major and -minor axes was done with SExtractor (Bertin \& Arnout 1996). Figure $7$ shows both the ellipticity distribution of all the galaxies in the UDF and the distribution of only the galaxies which were actually lensed into arcs. Based on a KS test, the two distributions differ at the 99.9 per cent confidence level. Clearly, the fraction of high ellipticity galaxies is higher in the population that is being lensed into arcs than the fraction in the whole UDF galaxy sample. A similar result has been found by Gao et al. (2009), who performed lensing simulations using galaxies from the Cosmic Evolution Survey (COSMOS) fields.  
\begin{figure}
\centering
\includegraphics[width=9cm]{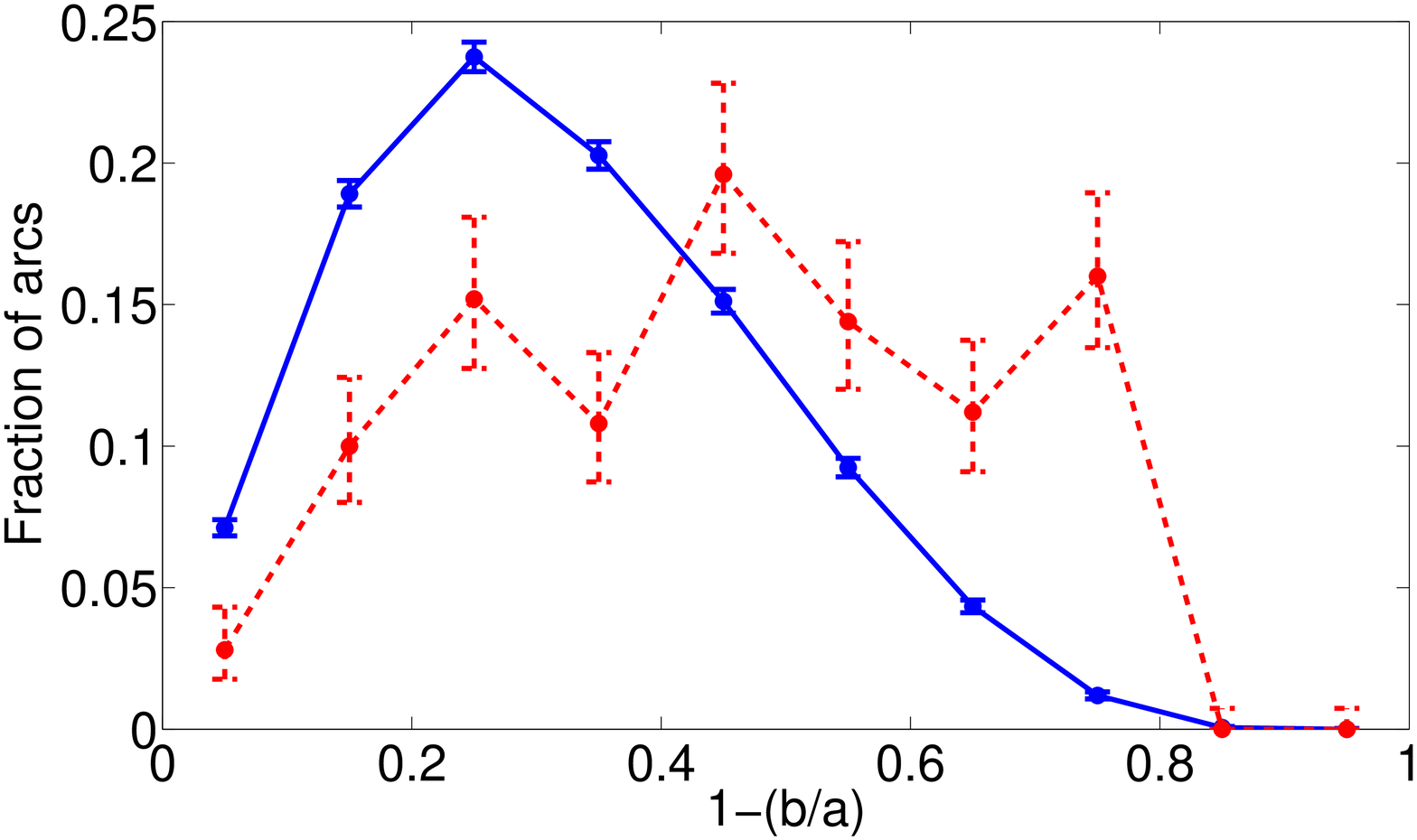}
\caption{Ellipticity distributions of all UDF galaxies (blue solid line) and of galaxies which are lensed into arcs (red dashed line).}
\end{figure}

\section{Comparison with observations}

Our new simulations of arc statistics at various redshifts can now be used for a comparison with the observed statistics. We compare our calculations
 both with the arc statistics of the MAssive Cluster Sample (MACS; Ebeling et al. 2001) at $z=0.4, 0.6$, that were presented in H10, and with the arc statistics for X-ray Brightest Abell-type Clusters of
  galaxies (XBACs) clusters at $z=0.2$ that were measured by H05.

MACS (Ebeling et al., 2001) has provided a statistically complete,
X-ray selected sample of the most X-ray luminous galaxy clusters at
$z>0.3$. Based on sources detected in the R\"ontgen Satellit (ROSAT)
All-Sky Survey (RASS, Voges et al. 1999), MACS covers 22,735 deg$^2$
of extragalactic sky ($|b|>20$ deg). The present MACS sample,
estimated to be at least 90\% complete, consists of 124 clusters, all of
which have optical spectroscopic redshifts. Owing to the high X-ray flux limit
of the RASS and the lower redshift limit of $z=0.3$, MACS clusters
have typical X-ray luminosities of $L_{X}\geq 5\times 10^{44}$ erg
s$^{-1}$ in the 0.1--2.4 keV band (Ebeling et al.\ 2007). MACS thus
probes the high end (${\rm M}_{200} \geq 10^{15}h^{-1}{\rm M}_{\sun}$) of the cluster mass function. The XBACs sample (Ebeling et al. 1996) that was analyzed by H05 spans a redshift range of $0.17< z <0.26$. The $0.1-2.4$ keV flux limit of $f_{X}\geq 5.0\times10^{-12}$ erg
cm$^{-2}$ s$^{-1}$ applied to this redshift range implies X-ray
luminosities of $L_{X}\geq4.1\times10^{44}$ erg s$^{-1}$, i.e.
similar to the MACS clusters at their higher redshifts. In terms of
survey volumes, MACS and XBACS probe volumes that are $220$ and $11$ times
larger, respectively,  than the Millenium simulation.

\subsection{Comparison of Millennium and MACS clusters}

\subsubsection{The $z=0.4$ subsample}

In H10, we analyzed subsamples of MACS clusters at 
  $0.3<z<0.5$ and $0.5<z<0.7$. 
We found the observed lensing efficiency of arcs with $l/w\geq 10$, at $0.3<z<0.5$, to be $0.74^{+0.23}_{-0.18}$. For comparison, we examine the efficiencies of the four Millennium clusters that have masses of ${\rm M}_{200}>10^{15}h^{-1}{\rm M}_{\sun}$, similar to the MACS cluster masses. The mean lensing efficiency for
 these Millennium clusters is $0.85^{+0.14}_{-0.12}$ arcs per cluster.
Similarly, for  $l/w\geq 8$, we observed 
$1.13^{+0.27}_{-0.22}$ arcs per MACS cluster (H10),
compared to $1.17^{+0.10}_{-0.09}$ arcs per Millennium cluster of this
 mass.
Thus, the mean lensing efficiency from our simulations 
is in excellent agreement
 with the MACS 
cluster lensing efficiency observed at these redshifts. 

Furthermore, a comparison of the observed and the predicted distributions of arc numbers in clusters (Fig. $4$, middle panel) reveals that the two distributions are remarkably similar. A Kolmogorov-Smirnov (KS) test on the simulated and observed distributions shows that the two are consistent with the null hypothesis of being drawn from the same parent distribution, with a high probability of $P=0.92$. The same null hypothesis, based on a comparison of
 the distributions of arc separations from cluster centers (Fig. 5) also has a high probablity, $P=0.22$. In terms of precision, 
the results for the $z\approx 0.4$ subsamples are the best, compared
 to the 
$z\approx 0.2$ and $z\approx 0.6$ results (to be discussed below), since
 the errors on both the observed and the predicted efficiencies are
 relatively small and comparable. We believe this is the 
strongest statistical evidence to date that the lensing efficiencies 
of observed galaxy 
clusters are in excellent agreement with numerical predictions
based on $\Lambda$CDM cosmology.

\subsubsection{The $z=0.6$ subsample}

 At $z=0.6$, when considering the full simulated subsample, 
with masses ${\rm M}_{200}\geq 7\times 10^{14} h^{-1} {\rm M}_{\sun}$, 
it appears
that the simulated arc production efficiency, $0.42^{+0.10}_{-0.08}$
$l/w>8$ arcs per 
Millennium cluster, is lower by a factor of
$3$ than observed in the $0.5<z<0.7$ MACS sample of H10,
$1.33^{+0.42}_{-0.33}$.
As seen in Fig. 4, the fraction of artificial clusters that do not
produce arcs is also higher by a factor of $3$ than the observed fraction. 
The null hypothesis that the arc-number distributions
 are derived from the same parent distribution
  has a probability of only $P=0.01$.
On the other hand, the simulated and observed distributions of 
arc separations from cluster centers at $z=0.6$ are consistent 
($P=0.24$ for the null hypothesis). 

However, as was the case for the $z=0.4$
sample, the full simulated sample is somewhat undermassive, compared
to the real MACS clusters.
Among the four Millennium clusters at $z=0.6$, 
two clusters (each with its 3 projections and 5 source realizations)
are responsible for the formation of $16$ arcs out of the total $18$ arcs  with $l/w\geq 10$ formed by all four clusters. The efficiency of only these two clusters is thus $0.53^{+0.17}_{-0.13}$ arcs per cluster, only a factor of $2$ lower than the observed efficiency of $1.08^{+0.39}_{-0.23}$ arcs per cluster. Furthermore, as shown in Figure $8$, one of the four clusters is in the process of formation and is still composed of several separate clumps. 
Only this one cluster, whose 15 projections$\times$realizations 
have an efficiency 
of $0.53^{+0.26}_{-0.18}$ arcs per cluster, has a mass above ${\rm M}_{200}\geq 10^{15} h^{-1} {\rm M}_{\sun}$,
 and its mass is only at the low-mass end of the range in the observed
 $z\sim 0.6$ MACS sample. 
\begin{figure}
\centering
\includegraphics[width=8cm]{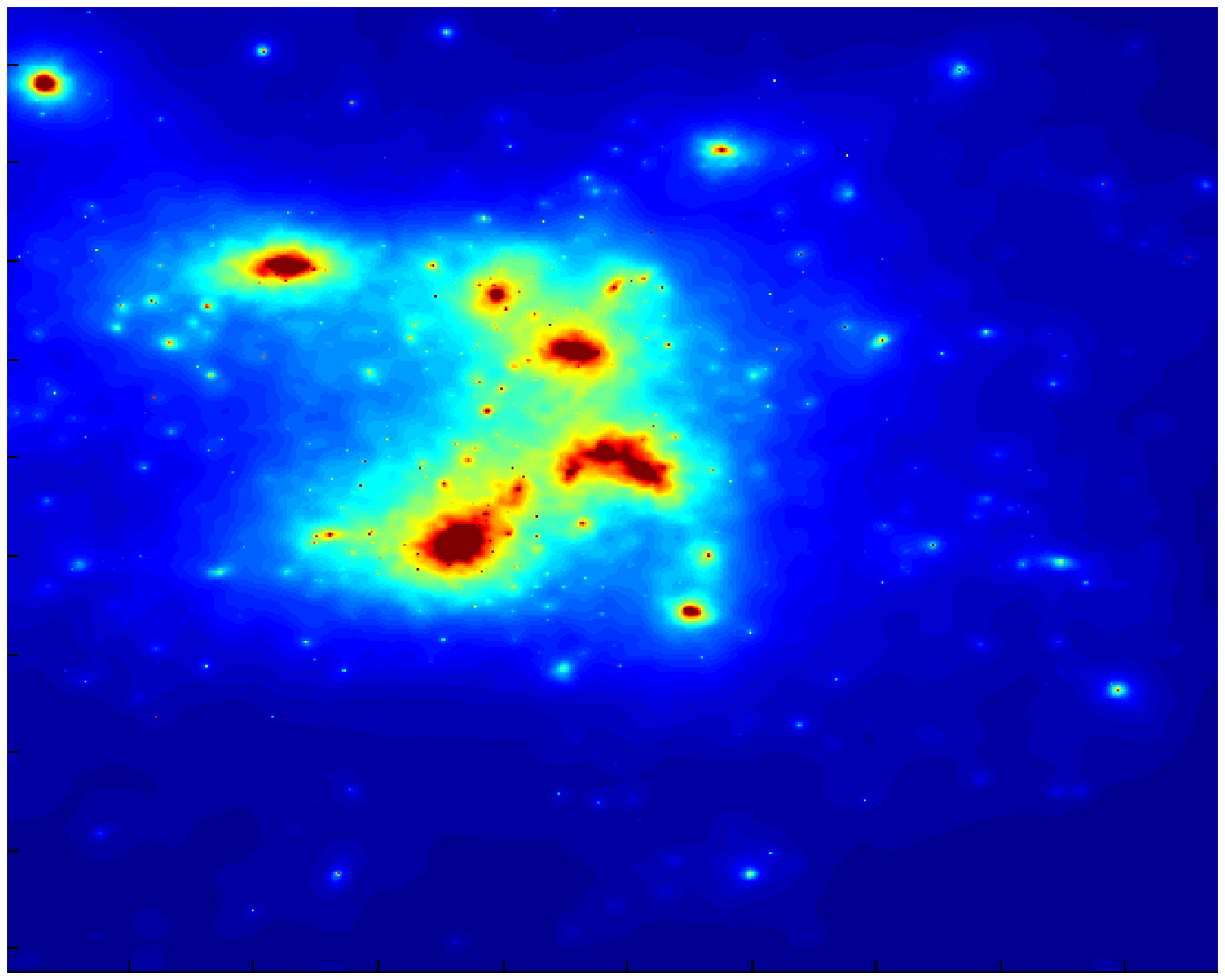}
\vskip 0.1cm
\includegraphics[width=8cm]{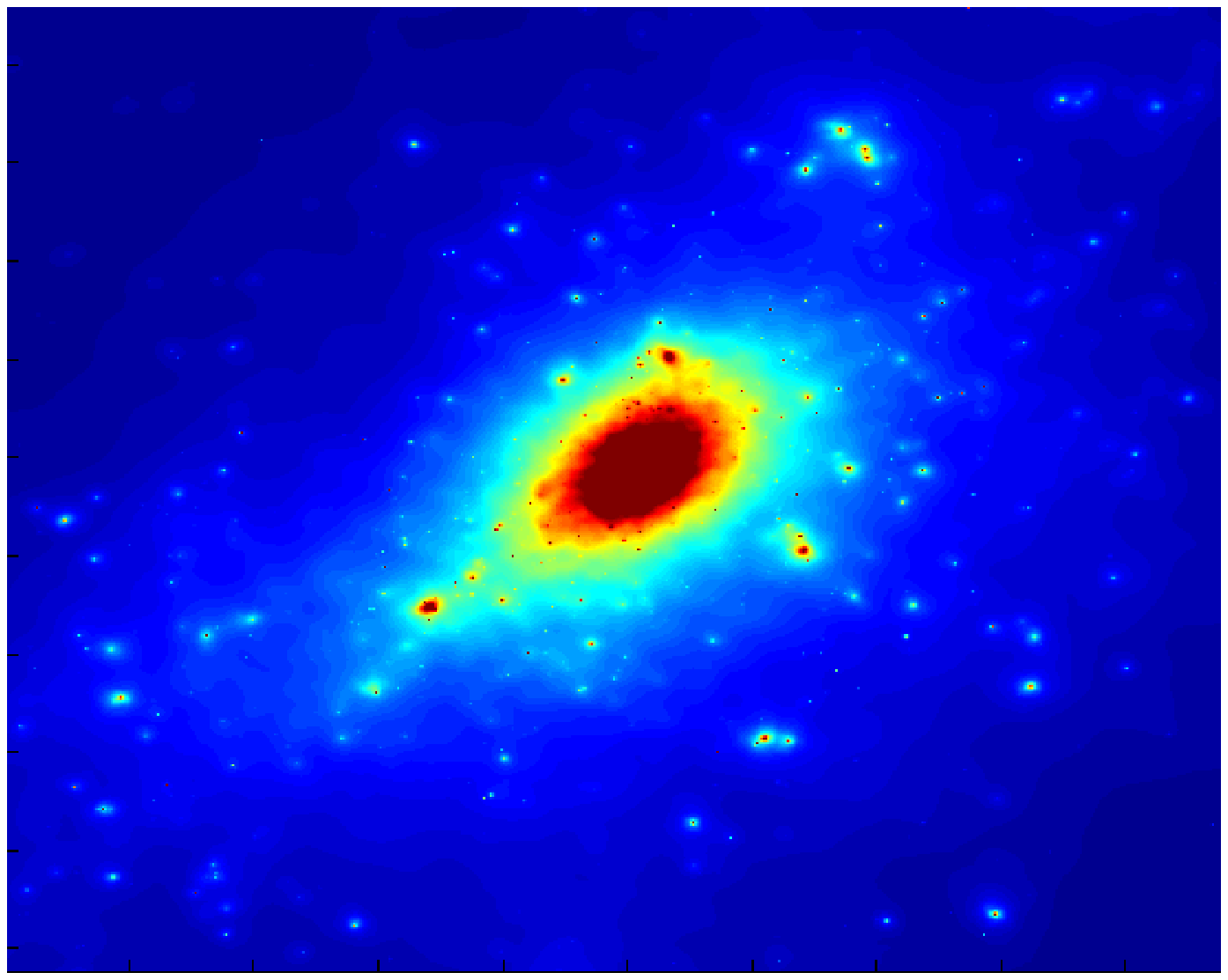}
\caption{A surface mass density map (in arbitrary units) of two of the
  four most massive clusters at $z=0.6$. As seen in the figure, one of
  the clusters (top panel) is composed of several separate clumps and
  is still in the process of formation, compared to the other cluster
  (bottom panel) that is more relaxed.}
\end{figure}
Clearly, we have too few clusters at $z=0.6$ with masses that are
similar to the MACS clusters for a proper comparison.  The question can be explored by N-body simulations with bigger box sizes, that
 allow the formation of a larger number of massive clusters at $z=0.6$.

\subsection{Comparison of Millennium and XBACs $z=0.2$ clusters}

In H05, we measured an observed efficiency of $1.2^{+0.46}_{-0.34}$ arcs ($l/w\geq10$) in a sample of $10$ XBACs clusters at $z=0.2$ (see also H10). 
The Smith et al. (2005) sample of clusters, analysed
by H05, spans a similar range of masses to those of our full $z=0.2$ 
Millennium sample, ${\rm M}_{200}\approx (7-20) \times
 10^{14} h^{-1} {\rm M}_{\sun}$.
The predicted lensing efficiency of the $14$ Millennium clusters that are in our sample at that redshift is $0.40^{+0.05}_{-0.04}$, again 
lower by a factor of 3 than the H05 observed result, a difference that is
significant at the $2.4\sigma$ level. 
In terms of the distributions, however, 
the null hypothesis that the simulated and observed distributions are drawn from the same parent distribution has high to moderate
KS probabilities: $P=0.33$ for the cluster arc number distributions,
 and $P=0.06$ for the distributions of arc separation from cluster centers.
Nevertheless, 
the $2.4\sigma$ difference between the integrated efficiencies again raises the B98 
question of whether the lensing properties of artificial clusters,
formed in a $\Lambda$CDM simulation at $z\approx 0.2$, are lower than
 their observed counterparts.

We first turn our attention again to the four most massive clusters in the Millennium simulations. These are the same four clusters that already comprise our cluster sample at $z=0.6$, with masses around $10^{15}h^{-1}{\rm M}_{\sun}$. 
At $z=0.4$, the lensing efficiency of these four clusters peaks at $0.85^{+0.14}
_{-0.12}$ arcs per cluster (for arcs with $l/w\geq 10$) and they are still efficient at $z=0.2$, with an efficiency of $0.68^{+0.12}_{-0.11}$ arcs per cluster. Therefore, if we choose to focus our comparison on only the four most massive clusters, we find a factor-of-2 enhancement in the lensing efficiency of the observed clusters, compared to the simulated clusters, but within the uncertainties the two are formally consistent
at the $1.5\sigma$ level.
We also note that in our simulation there are only seven clusters above our mass threshold of ${\rm M}_{200}\geq 7\times 10^{14}h^{-1}{\rm M}_{\sun}$ at $z=0.4$, and that their number doubles by $z=0.2$. Tracing back the mass evolution of these $14$ clusters from $z=0$ to $z=0.6$, we find that only eight of them gained more than $50 \%$ of their final mass above $z=0.6$. The average lensing efficiency of these latter clusters is $0.53^{+0.07}_{-0.07}$ which is $2.3$ times higher than the efficiency of the remaining $6$ clusters which formed later on. 
In \S6, below, we discuss some possible reasons behind the apparent discrepancy between
 the simulations 
and observations at $z=0.2$.

\section{Discussion and Summary}

We have presented a new set of realistic cluster lensing simulations,
and compared them to the observational arc statistics results of H10. 
In our simulations, we have lensed the UDF through artificial massive clusters from the Millennium N-body simulation. 
We have found that, at $z=0.4$, 
the most massive clusters, with ${\rm M}_{200}\geq 10^{15}h^{-1}{\rm M}_{\sun}$, are efficient lenses, 
producing an average of $0.85^{+0.14}_{-0.12}$ giant arcs with $l/w\geq 10$ per cluster. The results of our simulations are
 in excellent agreement with the observed efficiency of MACS clusters at $0.3<z<0.5$, presented in H10. 
At a higher redshift of $z=0.6$, we find that there are not enough 
massive clusters in the simulations in order to perform as strong a statistical comparison with the observations. Nevertheless, we do find two clusters that are already very efficient lenses at that redshift. 

We have further compared our 
new calculations of the lensing efficiency at $z=0.2$ with the observational 
results of H05. 
At this redshift, the lensing efficiency of our simulated cluster sample,
$0.40^{+0.05}_{-0.04}$ for arcs with $l/w\geq 10$, is low by a factor
of 3, compared to the observed efficiencies of H05. Thus, at cluster redshifts of $z=0.2$, the same redshift regime first investigated by B98, the discrepancy persists between the lensing efficiencies of real and simulated clusters, albeit at a lower level than found by B98. 

Given the agreement between the observed and the simulated samples at $z=0.4$,
an obvious candidate for explaining the discrepancy at $z=0.2$ is sample 
selection bias. The $0.3<z<0.5$ sample of MACS clusters analysed in H10
is an unbiased selection from a complete, X-ray-luminosity-selected, sample
by Ebeling et al. (2007). Targets were chosen by HST schedulers from the 
complete sample based only on scheduling convenience. 
The situation is less clear for the $z=0.2$
sample of Smith et al. (2005), analysed by H05. Smith et al. (2005) had
proposed to observe with HST (Program 8249, PI J.-P. Kneib) 16 clusters 
from a complete, luminosity-limited sample of 19
X-ray selected clusters, where the remaining 
three clusters already had suitable data in
the HST archive. In practice, however, time was granted to observe only
8 of 16 clusters requested. These 8 clusters, together with A2218 and
A2219 that already had archival data, 
constitute the Smith et al. (2005) sample of 10 XBACS clusters.
In choosing the 8 clusters to be observed,
the main criterion was distribution in RA to facilitate ground-based
followup, but there were also at least one or two clusters that were
excluded because they were thought to be unpromising lenses
(J.-P. Kneib, private communication). Thus, there was likely 
some degree of preselection favoring efficient lenses.

An additional possibility is that the observed H05 sample tends
to be more massive than the simulated Millennium $z=0.2$ sample. From Figure 1,
we can see that, among the 14 Millennium clusters at $z=0.2$,  8 (57\%)
are
in the mass range ${\rm M}_{200}=(7-9)\times 10^{14}h^{-1}{\rm
  M}_{\sun}$, 5 (36\%) are in the range ${\rm M}_{200}=(9-12)\times 10^{14}h^{-1}{\rm
  M}_{\sun}$, and one (7\%) with ${\rm M}_{200}>13\times 10^{15}h^{-1}{\rm
  M}_{\sun}$. By comparison, from Table 1 in H05 we can see that the
corresponding fractions in the observed Smith et al. (2005) sample are
40\%, 40\%, and 20\%, respectively. Although the masses listed in H05
are estimates based on the $L_X - M_{200}$ relation, which has
considerable scatter, this comparison suggests that the observed
sample may be more heavily weighted toward massive clusters than the 
simulated sample. This, in turn, could again contribute to the
discrepancy between the lensing efficiencies.

Alternatively or in parallel to these observational biases, 
some physical effect, not currently included in our simulations,
may set in at low redshifts, and make the real clusters  more 
efficient at arc production than the simulated clusters.
For example, Puchwein et al. (2005), Wambsganss et al. (2008), Rozo et al. (2008), and Mead et al. (2010) found that
the inclusion of intracluster gas in cosmological simulations, which of course is not included in the Millennium 
simulation, can under some conditions increase the lensing efficiency 
by up to a factor of a few. This increase in lensing efficiency is due to the steepening of the DM profile via adiabatic contraction (Gnedin et al. 2004). If this effect becomes important only at low redshifts,
it is of the right magnitude to explain our results. The cluster-formation 
physics behind X-ray selection is another possibility.
Torri et al. (2004) have found that
during a merger the strong-lensing cross section can be
enhanced by up to an order of magnitude. 
This is supported by Meneghetti et al. (2010), who find that 
SPH-simulation clusters that are strong lenses 
 tend to have higher X-ray luminosities 
than other clusters with the same mass. 
If the halo merger rate increases
at low redshifts, the low-$z$ sample could be affected in this way.
Indeed, as shown in \S4, when we limited our analysis to clusters that gained most of their
 mass early, at $z\geq 0.6$, the lensing efficiency increased. 
This suggests that there is some connection between the time a cluster 
forms and its lensing efficiency. This connection is not unexpected
since lensing efficiency depends also on the cluster concentration
which is correlated with the cluster formation
time (e.g. Wechsler et al. 2002).

In light of these results, 
we can now say that, in arc lensing statistics, there is at least one
redshift range, 
$z\approx 0.4$, for which there is a large and  unbiased observed
sample of clusters, there is a large artifical cluster sample with realistic
lensing simulations, and the lensing statistics of the two samples 
agree impressively.
At $z\approx 0.2$, there may be
problems at the factor-3 level, 
but there is unlikely to be an "order of magnitude problem", 
of the type raised by B98. At least some, if not all, of this
remaining discrepancy, is probably due to some combination of observed
sample pre-selection, cluster mass  mismatch between observations and
theory, and physical effects that are not included in the simulations,
such as the influence of ICM gas (Puchwein et al. 2005; Rozo et al. 2008).

The surface number density of lensed arcs 
over the whole sky depends on the product of 
the cluster mass distribution and the lensing efficiency as a function 
of cluster mass and redshift. Our study shows that the lensing
efficiency is probably quite close to the predictions of the
$\Lambda$CDM paradigm. It remains to be seen if so is the mass
function. The cluster mass function has long been considered a lithmus test
for cosmological models (e.g., Mortonson, Hu, \& Huterer 2010), and is
therefore at the focus of observational efforts using X-ray selection,
the red-sequence method, and the Sunyaev Zeldovich effect. Our results
concerning the lensing efficiencies of known clusters indicate that
future ``blind'' 
large-area lensed-arc surveys could provide another independent
avenue for measuring the cluster mass function.
For example,  Limousin et al. (2009) have carried
out such a blind search for arcs in
120 out of the 170 
square degrees of the
Canada-France-Hawaii Telescope Legacy Survey's Wide
Survey and found 13 galaxy-group scale lenses. The larger volume
needed to discover massive clusters via their strong lensing will be provided by the future Large Synoptic Survey Telescope (LSST).
Such cluster samples can provide an independent probe of the cluster mass function, with its great
diagnostic power for cosmological models.

\section*{Acknowledgements} 
We thank V. Springel for his help in accessing the Millennium data and J.-P. Kneib for discussions.
We thank the anonymous referee for constructive comments. AH acknowledges support by the Dan David Foundation.
SH acknowledges support by the Deutsche Forschungsgemeinschaft
within the Priority Programme 1177 under the project SCHN 342/6 and
by the German Federal Ministry of Education and Research (BMBF)
through the TR 33,  "The Dark Universe"'.
MB also acknowledges support by the BMBF, TR 33.

{}

\end{document}